# Modeling and Analysis of Dynamic Coupling Effects in a Free-Floating Space Manipulator

Erik Chumacero-Polanco
The University of Texas Rio Grande Valley
Brownsville, Texas, USA

## Abstract

Motion of a robotic manipulator mounted on a free-floating spacecraft induces base translation and rotation, creating a dynamic coupling that affects end-effector motion. This paper characterizes that coupling in a planar three-link space manipulator. Using momentum conservation, velocity mappings are developed between manipulator joints, the end effector, and the spacecraft base. Singular value decomposition is then used to examine coupling strength, directional anisotropy, and the orientation of the coupled motions. Numerical maps show that coupling strength is invariant to the first joint angle but depends strongly on the relative configuration of the remaining joints. Increasing the spacecraft base mass reduces joint-to-base and end-effector-to-base coupling, while its effect on the joint-to-end-effector mapping is smaller and more dependent on configuration. The translational end-effector-to-base mapping is highly anisotropic in some configurations: its base-velocity ellipse becomes increasingly flattened as the weak coupling mode approaches zero. The principal end-effector and base velocity directions also change their relative alignment across the configuration space, producing predominantly opposing responses in some regions and transverse or approximately codirectional responses in others. Together, these results provide a geometric description of how coupling magnitude and direction vary with manipulator configuration and spacecraft mass.

**Keywords:** Free-floating space manipulators; Dynamic coupling; Singular value decomposition; Coupling anisotropy; Coupling ellipsoids; Spacecraft base motion.

## 1. Introduction

The rapid development of space technology and the increasing complexity of orbital infrastructure are driving growing interest in robotic systems capable of performing demanding tasks in space [1,2]. Historically, many servicing, maintenance, and assembly operations have relied on astronauts operating in hazardous environments and at considerable mission cost [3,4]. At the same time, the increasing number of spacecraft and orbital objects has created a growing need for technologies capable of servicing existing assets and supporting future large-scale space infrastructure [5]. In-Space Servicing, Assembly, and Manufacturing (ISAM) using Space Manipulator Systems (SMSs) has consequently emerged as a promising approach for performing tasks such as transporting and assembling large orbital structures, servicing or capturing failed and decommissioned satellites, and supporting space-debris mitigation [6,7]. The feasibility of robotic servicing has been demonstrated through several missions and experimental programs, including the Shuttle Remote Manipulator System, the Robot Technology Experiment, Japan's Engineering Test Satellite VII, and the DARPA Orbital Express program [8–15].

A typical SMS consists of one or more robotic manipulators mounted on a spacecraft base. The manipulator is usually a fully actuated serial robot. Depending on how the spacecraft is controlled during manipulation, these systems are generally operated in free-flying or free-floating modes [16]. In free-flying operation, thrusters, reaction wheels, or other attitude-control devices actively regulate the spacecraft base. In free-floating operation, the base remains unactuated and is allowed to translate and rotate in response to manipulator motion. Manipulator motion generates reaction forces and torques on the spacecraft in both modes; however, these reactions can be actively compensated for in free-flying operation, whereas in free-floating operation they directly produce coupled base motion governed by conservation of linear and angular momentum [17]. This interaction between manipulator and spacecraft motion is commonly referred to as dynamic coupling. Because the resulting base translation and rotation modify the reference from which the manipulator operates, dynamic coupling can directly affect end-effector (EE) positioning and tracking accuracy [18]. Dynamic coupling between manipulator and spacecraft in free-floating robotic manipulators (FFRMs) is analyzed in the present study.

A substantial body of research has therefore focused on compensating for or suppressing the effects of dynamic coupling in FFRMs [18]. Proposed approaches include sliding-mode, optimal, robust, adaptive, neural-network, and fuzzy control methods, as well as optimization-based trajectory generation [19–31]. These methods have demonstrated the ability to improve EE tracking, regulate spacecraft attitude, accommodate model uncertainty, and satisfy operational constraints. In all these approaches, the spacecraft reaction generated by manipulator motion is treated primarily as a disturbance to be minimized or compensated.

An alternative perspective is to regard dynamic coupling as an intrinsic system property that can be characterized and, when appropriate, deliberately exploited**.** Early studies demonstrated that the dynamic coupling generated by momentum conservation could be exploited to accomplish spacecraft reorientation through coordinated manipulator motion [32]. Xu (1993) [33] , for example, investigated measures of dynamic coupling between EE and spacecraft motion and their implications for motion planning and system design. Nenchev et al. subsequently introduced the Reaction Null Space concept [34], exploiting the structure of the coupling relationship to identify manipulator motions that generate no reaction at the spacecraft base. Coupling information has also been incorporated into optimization-based trajectory generation; Misra et al. (2017) [23] , for example, considered spacecraft disturbance within a constrained trajectory-planning framework. Jia et al. (2019) [22] demonstrated another use of dynamic coupling by employing active-joint motion to indirectly regulate passive joints following free-swinging joint failures. More recently, a previous study from the author [35]  examined the influence of manipulator configuration and inertia distribution on base reaction, further illustrating the fundamental dependence of spacecraft response on the coupled dynamics of the complete system.

In parallel with these application-oriented developments, several studies have sought to quantitatively characterize dynamic coupling itself. Zhou et al. (2019) [36] introduced singular-value-based dynamic coupling factors and coupling ellipsoids to characterize coupling magnitude and its geometric properties and investigated their dependence on system parameters and manipulator configuration. Peng et al. (2020) [37] extended quantitative coupling analysis to dual-arm free-floating space robotic systems by deriving multiple coupling relationships between the two manipulators and the spacecraft base and defining dynamic coupling factors to evaluate these interactions. Their analysis demonstrated the dependence of the coupling factors on manipulator configuration, mass and inertial properties, payload, and manipulator installation geometry, and further illustrated their use for reducing spacecraft disturbance through trajectory planning. More recently, Das et al. (2025) [32] employed singular value decomposition (SVD) to identify dominant coupling directions and incorporated coupling information into a trajectory-optimization framework aimed at utilizing, rather than simply suppressing, favorable dynamic interactions. Together, these studies establish coupling factors, singular values, and principal coupling directions as useful tools for understanding the configuration-dependent interaction between space manipulators and their free-floating bases.

Despite these advances, the maximum coupling magnitude alone does not completely characterize how motion is transmitted between the EE and spacecraft base. Two configurations with comparable maximum coupling may differ substantially in the relative gains of their strong and weak coupling modes. Similarly, knowledge of the principal coupling directions does not by itself indicate the relative orientation between the imposed EE motion and the resulting base response. Recent SVD-based coupling analyses have demonstrated the importance of considering the distribution of coupling strengths and whether coupling-induced motion assists a desired task direction [32]. However, this task-oriented measure differs from directly characterizing the relative orientation between the principal EE input directions and their corresponding induced base-motion directions, as well as how this relationship evolves throughout the manipulator configuration space. While previous studies have examined coupling magnitude, coupling ellipsoids, principal coupling directions, and configuration-dependent coupling characteristics [32,33,36,37], the combined relationship among coupling strength, strong-to-weak anisotropy, and EE–base directional alignment has received comparatively limited attention. A systematic characterization of these complementary properties can therefore provide a more complete description of not only how strongly EE motion affects the spacecraft base, but also how selectively that motion is transmitted with respect to direction and where the resulting base response is directed.

A related limitation of a global EE-to-base coupling measure is that it combines translational and rotational motion and therefore does not explicitly reveal the physical nature of the spacecraft response. EE translation can induce both base translation and rotation, while EE rotation can similarly produce translational and rotational base responses. Distinguishing these individual motion-transfer relationships is important because a manipulator configuration that attenuates one component of the spacecraft response does not necessarily suppress another. A more complete characterization therefore requires examining not only the global EE-to-base interaction but also the individual translation-to-translation, translation-to-rotation, rotation-to-translation, and rotation-to-rotation coupling mechanisms.

To address these issues, this work presents an SVD-based characterization of dynamic coupling in a planar free-floating robotic manipulator based on three complementary properties: coupling strength, coupling anisotropy, and directional alignment. First, the manipulator-to-base, manipulator-to-EE, and EE-to-base coupling relationships are evaluated throughout the manipulator configuration space, including the effects of spacecraft base mass and the rotational symmetry of the planar system. The EE-to-base interaction is then resolved into translation-to-translation, translation-to-rotation, rotation-to-translation, and rotation-to-rotation relationships to identify the different mechanisms through which EE motion is transmitted to the spacecraft. The strong and weak modes of the translational EE-to-base coupling are subsequently interpreted geometrically through velocity-coupling ellipses, including highly anisotropic and rank-deficient configurations. Finally, the relative alignment between the corresponding EE and base velocity directions is quantified throughout the configuration space to distinguish codirectional, transverse, and opposing coupling behaviors. Together, these analyses provide a unified characterization of how strongly motion is transmitted, how directionally selective that transmission is, and where the resulting coupled motion is directed, providing a foundation for future coupling-aware configuration selection, motion planning, and control of free-floating space robotic systems.

The remainder of this paper is organized as follows. Section 2 introduces the planar three-link FFRM model and develops the kinematic and dynamic equations used in the analysis. Section 3 formulates the dynamic coupling relationships and introduces SVD-based metrics. Section 4 presents and discusses the numerical results, examining the effects of manipulator configuration and spacecraft base mass on the coupling characteristics. Finally, Section 5 summarizes the main findings and outlines directions for future work.

## 2. Kinematic and Dynamic Model

A free-floating robotic manipulator (FFRM) consists of a spacecraft base and a robotic manipulator mounted on it. The base is free to translate and rotate in response to manipulator motion, while the manipulator is modeled as a fully actuated serial robot. To facilitate a systematic analysis of the dynamic coupling while retaining the fundamental interaction between manipulator and base motion, the six-degree-of-freedom (6-DOF) planar FFRM shown in Fig. 1 is considered. The system comprises three unactuated base coordinates and three actuated revolute joints. The spacecraft base translates and rotates in the inertial plane $\{X_I, Y_I\}$, while the three-link manipulator is actuated by the joint torques $\tau_1, \tau_2, \tau_3$. The geometric and inertial parameters include the link lengths $l_i$, link center-of-mass locations $l_{ci}$, base and link masses $m_0$and $m_i$, and the corresponding moments of inertia $I_0$ and $I_i$, respectively, for $i = 1,2,3$.

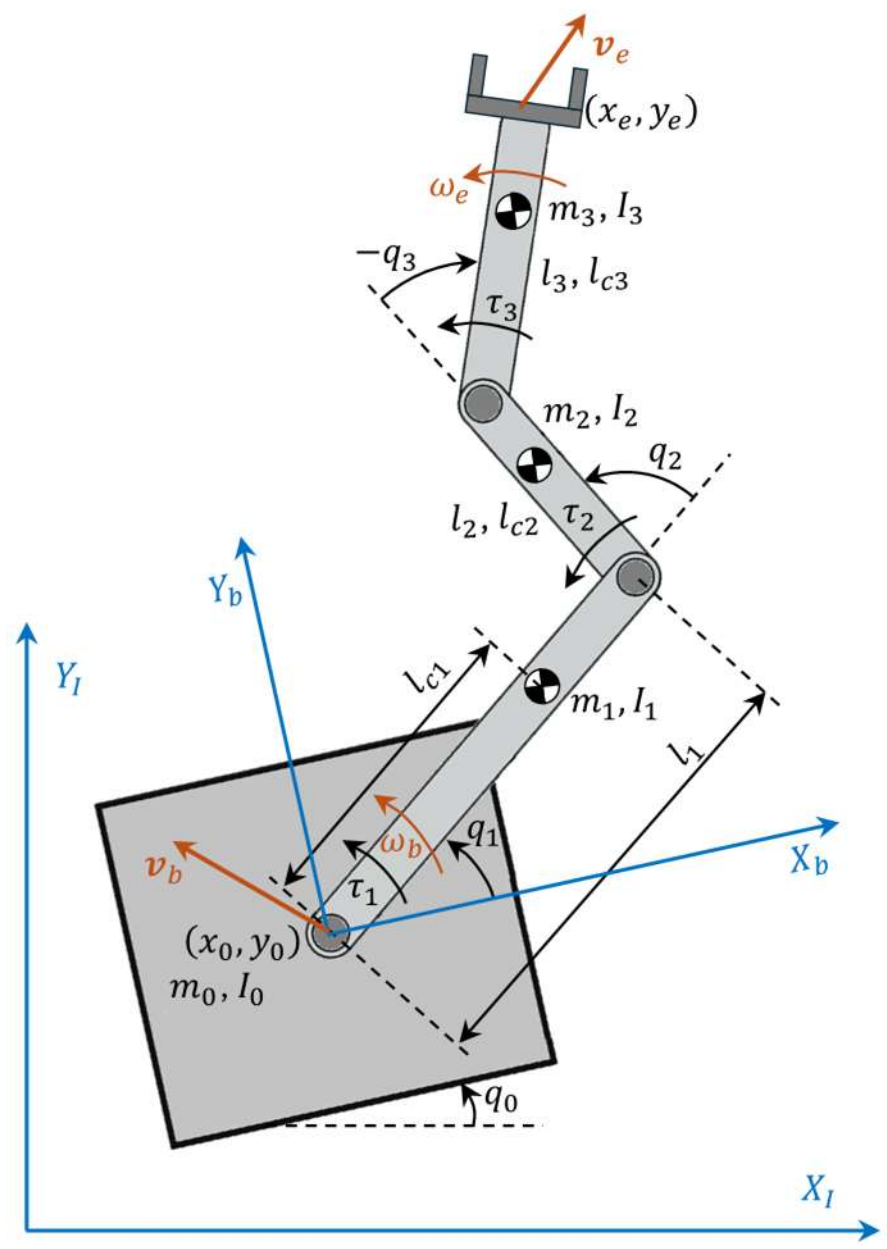

*Figure 1: Planar three-link FFRM model and coordinate definitions, showing the inertial and spacecraft-fixed reference frames, base and joint angles, link lengths and center-of-mass locations, inertial parameters, and joint torques.*

The generalized-coordinate vector is defined as

$$\boldsymbol{q} = [x_0 \quad y_0 \quad q_0 \quad q_1 \quad q_2 \quad q_3]^T = [\boldsymbol{q}_b^T \quad \boldsymbol{q}_a^T]^T,$$

where $x_0$ and $y_0$ define the position of the spacecraft-base center of mass (COM) in the inertial frame, $q_0$ defines the base orientation, and $q_i, q_2, q_3$ are the local manipulator joint angles. Accordingly, position vectors

$$\boldsymbol{q}_b = [x_0 \quad y_0 \quad q_0]^T \text{ and } \boldsymbol{q}_a = [q_1 \quad q_2 \quad q_3]^T,$$

denote the base and manipulator/arm generalized coordinates, respectively.

The spacecraft base and manipulator links are modeled as rigid bodies with lumped mass and inertia properties. The system is assumed to operate in a microgravity environment with negligible external forces and torques. The total linear and angular momenta are therefore conserved. Unless otherwise stated, the system is assumed to be initially at rest, such that its initial total momentum is zero.

## 2.2 Kinematic Model

The kinematic model of the FFRM shown in Figure 1 defines the positions of the joints and center of mass (COM), as well as the pose of the end effector (EE), with respect to the inertial frame. For the system considered, the base ($\boldsymbol{r}_0$), joint ($\boldsymbol{r}_1, \boldsymbol{r}_2$), link COM ($\boldsymbol{r}_{c1}, \boldsymbol{r}_{c2}, \boldsymbol{r}_{c3}$), and EE ($\boldsymbol{r}_e$) positions are defined as

$$\boldsymbol{r}_0 = [x_0 \quad y_0]^T, \qquad \boldsymbol{r}_{c1} = \boldsymbol{r}_0 + l_{c1}\boldsymbol{u}_{01},$$
$$\boldsymbol{r}_1 = \boldsymbol{r}_0 + l_1\boldsymbol{u}_{01}, \qquad \boldsymbol{r}_{c2} = \boldsymbol{r}_1 + l_{c2}\boldsymbol{u}_{02},$$
$$\boldsymbol{r}_2 = \boldsymbol{r}_1 + l_2\boldsymbol{u}_{02}, \qquad \boldsymbol{r}_{c3} = \boldsymbol{r}_2 + l_{c3}\boldsymbol{u}_{03},$$

and

$\boldsymbol{r}_e = \boldsymbol{r_2} + l_3\boldsymbol{u}_{03};$

where $\boldsymbol{u}_{0j} = \left[\cos(q_{0j}) \quad \sin(q_{0j})\right]^T$ denotes the unit vector along link $j$, respectively, for $j = 1,2,3$. In addition, $q_{0j} = \sum_{i=0}^{j} q_i$ and $\boldsymbol{H} = [0 \quad -1; 1 \quad 0]$ is the $90°$rotation matrix.

The linear velocities of the spacecraft base ($\boldsymbol{v}_0$), manipulator joints ($\boldsymbol{v}_1, \boldsymbol{v}_2$), links' COMs ($\boldsymbol{v}_{c0}, \boldsymbol{v}_{c1}, \boldsymbol{v}_{c2}, \boldsymbol{v}_{c3}$), and EE ($\boldsymbol{v}_e$) are obtained by differentiating their position vectors with respect to time, which yields to

$\boldsymbol{v}_0 = \boldsymbol{v}_{c0} = [\dot{x}_0 \quad \dot{y}_0]^T,$

$\boldsymbol{v}_1 = \boldsymbol{v}_0 + l_1\boldsymbol{H}\boldsymbol{u}_{01}\dot{q}_{01},$

$\boldsymbol{v}_2 = \boldsymbol{v}_0 + l_1\boldsymbol{H}\boldsymbol{u}_{01}\dot{q}_{01} + l_2\boldsymbol{H}\boldsymbol{u}_{02}\dot{q}_{02},$

$\boldsymbol{v}_{c1} = \boldsymbol{v}_{c0} + l_{c1}\boldsymbol{H}\boldsymbol{u}_{01}\dot{q}_{01},$

$\boldsymbol{v}_{c2} = \boldsymbol{v}_{c0} + l_1\boldsymbol{H}\boldsymbol{u}_{01}\dot{q}_{01} + l_{c2}\boldsymbol{H}\boldsymbol{u}_{02}\dot{q}_{02},$

$\boldsymbol{v}_{c3} = \boldsymbol{v}_{c0} + l_1\boldsymbol{H}\boldsymbol{u}_{01}\dot{q}_{01} + l_2\boldsymbol{H}\boldsymbol{u}_{02}\dot{q}_{02} + l_{c3}\boldsymbol{H}\boldsymbol{u}_{03}\dot{q}_{03}$, and

$\boldsymbol{v}_e = \boldsymbol{v}_0 + l_1\boldsymbol{H}\boldsymbol{u}_{01}\dot{q}_{01} + l_2\boldsymbol{H}\boldsymbol{u}_{02}\dot{q}_{02} + l_3\boldsymbol{H}\boldsymbol{u}_{03}\dot{q}_{03},$

The preceding velocity expressions are written using Jacobian matrices, where the generalized coordinates vector $\boldsymbol{q} = [\dot{x}_0 \ \dot{y}_0 \ \dot{q}_0 \ \dot{q}_1 \ \dot{q}_2 \ \dot{q}_3\ ]^T$is written as $\boldsymbol{q} = \left[\boldsymbol{q}_b^T \quad \boldsymbol{q}_a^T\right]^T$; with $\boldsymbol{q}_b = [\dot{x}_0 \ \dot{y}_0 \ \dot{q}_0]^T$ and $\boldsymbol{q}_a = [\dot{q}_1 \ \dot{q}_2 \ \dot{q}_3]^T$**.** That is, base's COM velocity is written as

$$\boldsymbol{v}_{c0} = \begin{bmatrix} 1 & 0 & 0 & 0 & 0 & 0 \\ 0 & 1 & 0 & 0 & 0 & 0 \end{bmatrix} \begin{bmatrix} \dot{x}_0 \\ \dot{y}_0 \\ \dot{q}_0 \\ \dot{q}_1 \\ \dot{q}_2 \\ \dot{q}_3 \end{bmatrix} = \left[\begin{bmatrix} 1 & 0 & 0 \\ 0 & 1 & 0 \end{bmatrix} \quad \begin{bmatrix} 0 & 0 & 0 \\ 0 & 0 & 0 \end{bmatrix}\right] \begin{bmatrix} \dot{\boldsymbol{q}}_b \\ \dot{\boldsymbol{q}}_a \end{bmatrix} = \boldsymbol{J}_{v0}\dot{\boldsymbol{q}},$$

where Jacobian matrix $\boldsymbol{J}_{v0}$ has been introduced. Similarly, the COM velocity vectors of links $1$ through $3$ are writte as

$$\boldsymbol{v}_{c1} = \left[\left[\boldsymbol{I}_{2\times2} \quad l_{c1}\boldsymbol{H}\boldsymbol{u}_{01}(\boldsymbol{q})\right] \quad \left[l_{c1}\boldsymbol{H}\boldsymbol{u}_{01}(\boldsymbol{q}) \begin{bmatrix} 1 \\ 0 \\ 0 \end{bmatrix}^T\right]\right] \begin{bmatrix} \dot{\boldsymbol{q}}_b \\ \dot{\boldsymbol{q}}_a \end{bmatrix} = \boldsymbol{J}_{v1}(\boldsymbol{q})\dot{\boldsymbol{q}},$$

$$\boldsymbol{v}_{c2} = \left(\left[\boldsymbol{I}_{2\times2} \quad l_1\boldsymbol{H}\boldsymbol{u}_{01} + l_{c2}\boldsymbol{H}\boldsymbol{u}_{02}\right] \quad \left[l_1\boldsymbol{H}\boldsymbol{u}_{01} \begin{bmatrix} 1 \\ 0 \\ 0 \end{bmatrix}^T + l_{c2}\boldsymbol{H}\boldsymbol{u}_{02} \begin{bmatrix} 1 \\ 1 \\ 0 \end{bmatrix}^T\right]\right) \begin{bmatrix} \dot{\boldsymbol{q}}_b \\ \dot{\boldsymbol{q}}_a \end{bmatrix} = \boldsymbol{J}_{v2}(\boldsymbol{q})\dot{\boldsymbol{q}}, \text{ and}$$

$$\boldsymbol{v}_{c3} = \left(\left[\boldsymbol{I}_{2\times2} \quad l_1\boldsymbol{H}\boldsymbol{u}_{01} + l_2\boldsymbol{H}\boldsymbol{u}_{02} + l_{c3}\boldsymbol{H}\boldsymbol{u}_{03}\right] \quad \left[l_1\boldsymbol{H}\boldsymbol{u}_{01} \begin{bmatrix} 1 \\ 0 \\ 0 \end{bmatrix}^T + l_2\boldsymbol{H}\boldsymbol{u}_{02} \begin{bmatrix} 1 \\ 1 \\ 0 \end{bmatrix}^T + l_{c3}\boldsymbol{H}\boldsymbol{u}_{03} \begin{bmatrix} 1 \\ 1 \\ 1 \end{bmatrix}^T\right]\right) \begin{bmatrix} \dot{\boldsymbol{q}}_b \\ \dot{\boldsymbol{q}}_a \end{bmatrix} = \boldsymbol{J}_{v3}(\boldsymbol{q})\dot{\boldsymbol{q}},$$

where Jacobian matrices $\boldsymbol{J}_{vi}(\boldsymbol{q})$, $j = 1,2,3$ have been introduced.
As for the EE velocity vector, similar procedure yields

$$\boldsymbol{v}_e = \left( \underbrace{\begin{bmatrix} \boldsymbol{I}_{2\times 2} & l_1\boldsymbol{Hu}_{01} + l_2\boldsymbol{Hu}_{02} + l_3\boldsymbol{Hu}_{03} \end{bmatrix}}_{\boldsymbol{J}_{ve,b}\in\mathbb{R}^{2x3}} \quad \underbrace{\begin{bmatrix} l_1\boldsymbol{Hu}_{01}\begin{bmatrix}1\\0\\0\end{bmatrix}^T + l_2\boldsymbol{Hu}_{02}\begin{bmatrix}1\\1\\0\end{bmatrix}^T + l_3\boldsymbol{Hu}_{03}\begin{bmatrix}1\\1\\1\end{bmatrix}^T \end{bmatrix}}_{\boldsymbol{J}_{ve,a}\in\mathbb{R}^{2x3}} \right) \begin{bmatrix}\dot{\boldsymbol{q}}_b\\ \dot{\boldsymbol{q}}_a\end{bmatrix} = \boldsymbol{J}_{ve}(\boldsymbol{q})\dot{\boldsymbol{q}}, \tag{1}$$

The Jacobian matrices $\boldsymbol{J}_{vi}(\boldsymbol{q}) \in \mathbb{R}^{2\times 6}$, $i = 0,1,2,3$, can be partitioned into $2\times 3$ base and manipulator components. Following the same partition for the EE Jacobian gives

$$\boldsymbol{v}_e = \boldsymbol{J}_{ve,b}\dot{\boldsymbol{q}}_b + \boldsymbol{J}_{ve,a}\dot{\boldsymbol{q}}_a. \tag{2}$$

To include both translational and rotational EE motion, the EE velocity vector is defined as

$$\dot{\boldsymbol{q}}_e = \begin{bmatrix}\boldsymbol{v}_e\\ \omega_e\end{bmatrix} = \begin{bmatrix}\boldsymbol{J}_{ve,b} & \boldsymbol{J}_{ve,a}\\ [0\ 0\ 1] & [1\ 1\ 1]\end{bmatrix}\begin{bmatrix}\dot{\boldsymbol{q}}_b\\ \dot{\boldsymbol{q}}_a\end{bmatrix} = [\boldsymbol{J}_{eb} \quad \boldsymbol{J}_{ea}]\begin{bmatrix}\dot{\boldsymbol{q}}_b\\ \dot{\boldsymbol{q}}_a\end{bmatrix} = \boldsymbol{J}_e(\boldsymbol{q})\dot{\boldsymbol{q}}, \tag{3}$$

where

$$\boldsymbol{J}_{eb} = \begin{bmatrix}\boldsymbol{J}_{ve,b}\\ [0\ 0\ 1]\end{bmatrix} \quad \text{and} \quad \boldsymbol{J}_{ea} = \begin{bmatrix}\boldsymbol{J}_{ve,a}\\ [1\ 1\ 1]\end{bmatrix}.$$

Finally, the angular velocities of the spacecraft base, manipulator links, and EE are expressed in terms of the generalized velocities as

$$\boldsymbol{\omega}_0 = \begin{bmatrix}0\\0\\ \dot{q}_0\end{bmatrix} = \begin{bmatrix}0&0&0&0&0&0\\0&0&0&0&0&0\\0&0&1&0&0&0\end{bmatrix}\begin{bmatrix}\dot{x}_0\\ \dot{y}_0\\ \dot{q}_0\\ \dot{q}_1\\ \dot{q}_2\\ \dot{q}_3\end{bmatrix} = \boldsymbol{J}_{\omega 0}\dot{\boldsymbol{q}},$$

$$\boldsymbol{\omega}_1 = \begin{bmatrix}0\\0\\ \dot{q}_0 + \dot{q}_1\end{bmatrix} = \begin{bmatrix}0&0&0&0&0&0\\0&0&0&0&0&0\\0&0&1&1&0&0\end{bmatrix}\begin{bmatrix}\dot{x}_0\\ \dot{y}_0\\ \dot{q}_0\\ \dot{q}_1\\ \dot{q}_2\\ \dot{q}_3\end{bmatrix} = \boldsymbol{J}_{\omega 1}\dot{\boldsymbol{q}},$$

$$\boldsymbol{\omega}_2 = \begin{bmatrix}0\\0\\ \dot{q}_0 + \dot{q}_1 + \dot{q}_2\end{bmatrix} = \begin{bmatrix}0&0&0&0&0&0\\0&0&0&0&0&0\\0&0&1&1&1&0\end{bmatrix}\begin{bmatrix}\dot{x}_0\\ \dot{y}_0\\ \dot{q}_0\\ \dot{q}_1\\ \dot{q}_2\\ \dot{q}_3\end{bmatrix} = \boldsymbol{J}_{\omega 2}\dot{\boldsymbol{q}}, \text{ and}$$

$$\boldsymbol{\omega}_3 = \boldsymbol{\omega}_e = \begin{bmatrix} 0 \\ 0 \\ \dot{q}_0 + \dot{q}_1 + \dot{q}_2 + \dot{q}_3 \end{bmatrix} = \begin{bmatrix} 0 & 0 & 0 & 0 & 0 & 0 \\ 0 & 0 & 0 & 0 & 0 & 0 \\ 0 & 0 & 1 & 1 & 1 & 1 \end{bmatrix} \begin{bmatrix} \dot{x}_0 \\ \dot{y}_0 \\ \dot{q}_0 \\ \dot{q}_1 \\ \dot{q}_2 \\ \dot{q}_3 \end{bmatrix} = \boldsymbol{J}_{\omega 3}\dot{\boldsymbol{q}}.$$

The linear and angular Jacobian matrices derived above describe the velocity kinematics of the planar FFRM and are used to formulate the kinetic energy and equations of motion of the system.

### 2.3 Dynamic Model

The dynamic model of the FFRM is adopted from established Lagrangian formulations. In the absence of gravity, the Lagrangian of the system shown in Figure 1 is

$$L(\boldsymbol{q}, \dot{\boldsymbol{q}}) = \frac{1}{2}\dot{\boldsymbol{q}}\boldsymbol{M}(\boldsymbol{q})\dot{\boldsymbol{q}},$$

where $\boldsymbol{M}(\boldsymbol{q})$ stands for the inertia matrix and is partitioned as $\boldsymbol{M}(\boldsymbol{q}) = \boldsymbol{M}_v(\boldsymbol{q}) + \boldsymbol{M}_\omega(\boldsymbol{q})$. Matrices $\boldsymbol{M}_v(\boldsymbol{q})$ and $\boldsymbol{M}_\omega(\boldsymbol{q})$ denote translational and rotational components and are defined as

$\boldsymbol{M}_v(\boldsymbol{q}) = \sum_{i=0}^{3} m_i \boldsymbol{J}_{vi}^T(\boldsymbol{q})\boldsymbol{J}_{vi}(\boldsymbol{q})$ and

$\boldsymbol{M}_\omega(\boldsymbol{q}) = \sum_{i=0}^{3} I_i \boldsymbol{J}_{\omega i}^T(\boldsymbol{q})\boldsymbol{J}_{\omega i}(\boldsymbol{q})$.

Under this notation, the equation of motion of the FFRM can be written in a compact form as

$$\boldsymbol{M}(\boldsymbol{q})\ddot{\boldsymbol{q}} + \boldsymbol{h}(\boldsymbol{q}, \dot{\boldsymbol{q}}) = \boldsymbol{\tau},$$

where $\ddot{\boldsymbol{q}}$ denotes generalized acceleration vector and $\boldsymbol{h}(\boldsymbol{q}, \dot{\boldsymbol{q}})$ is the vector of nonlinear dynamic terms defined as

$\boldsymbol{h} = \dot{\boldsymbol{M}}\dot{\boldsymbol{q}} - \frac{1}{2}\left[\frac{\partial}{\partial \boldsymbol{q}}(\dot{\boldsymbol{q}}^T M \dot{\boldsymbol{q}})\right]^T$, and $\boldsymbol{\tau}$ denotes the vector of generalized forces.

Finally, $\boldsymbol{M}(\boldsymbol{q})$, $\boldsymbol{h}(\boldsymbol{q}, \dot{\boldsymbol{q}})$, $\boldsymbol{\tau}$, and $\boldsymbol{q}$ are split into base and manipulator components as

$$\begin{bmatrix} \boldsymbol{M}_b(\boldsymbol{q}) & \boldsymbol{M}_{ab}(\boldsymbol{q}) \\ \boldsymbol{M}_{ab}^T(\boldsymbol{q}) & \boldsymbol{M}_a(\boldsymbol{q}) \end{bmatrix} \begin{bmatrix} \ddot{\boldsymbol{q}}_b \\ \ddot{\boldsymbol{q}}_a \end{bmatrix} + \begin{bmatrix} \boldsymbol{h}_b(\boldsymbol{q}, \dot{\boldsymbol{q}}) \\ \boldsymbol{h}_a(\boldsymbol{q}, \dot{\boldsymbol{q}}) \end{bmatrix} = \begin{bmatrix} \boldsymbol{\tau}_b \\ \boldsymbol{\tau}_a \end{bmatrix},$$

where $\boldsymbol{\tau}_b = [F_x \quad F_y \quad \tau_0]^T = [0\ 0\ 0]$ denotes wrench vector acting at the base and $\boldsymbol{\tau}_a = [\tau_1 \quad \tau_2 \quad \tau_3]^T$ stands for the joint torques applied on the manipulator.

## 3. Coupling Formulation and Metrics

This section develops the velocity-level dynamic coupling relationships used to characterize the FFRM. The coupling between manipulator motion, spacecraft-base motion, and end-effector (EE) motion is first formulated from momentum conservation and differential kinematics. A direct EE-to-base coupling relationship is subsequently obtained, together with individual mappings describing the translational and rotational components of the coupled motion. Singular value decomposition (SVD) is then used to characterize three complementary properties of the dynamic coupling: its strength, directional anisotropy, and the alignment between the corresponding input and output velocity directions.

### 3.1 Manipulator-Base coupling

In this study, gravity and external forces/torques acting on the system are neglected. Consequently, the motion of the system is generated exclusively by the internal forces and moments produced by the manipulator actuators. Under these assumptions, the total linear and angular momenta of the system is preserved throughout the motion. The generalized momentum associated with the base coordinates is obtained from the Lagrangian as

$$\boldsymbol{p}_b = \frac{\partial L(\boldsymbol{q})}{\partial \dot{\boldsymbol{q}}_b} = \boldsymbol{M}_b(\boldsymbol{q})\dot{\boldsymbol{q}}_b + \boldsymbol{M}_{ab}(\boldsymbol{q})\dot{\boldsymbol{q}}_a,$$

where vector $\boldsymbol{p}_b = [p_x \quad p_y \quad h_z]^T$ contains the total FFRM COM linear momentum along $x$ and $y$ directions ($p_x$ and $p_y$) and angular momentum ($h_z$) about an axis passing through the FFRM's COM. Assuming that the system is initially at rest, the initial momentum is zero

$$\boldsymbol{M}_b(\boldsymbol{q})\dot{\boldsymbol{q}}_b + \boldsymbol{M}_{ab}(\boldsymbol{q})\dot{\boldsymbol{q}}_a = \boldsymbol{0}.$$

Solving for the base velocity yields

$$\dot{\boldsymbol{q}}_b = \boldsymbol{C}_{ba}(\boldsymbol{q})\dot{\boldsymbol{q}}_a, \tag{4}$$

where

$$\boldsymbol{C}_{ba}(\boldsymbol{q}) = -\boldsymbol{M}_b^{-1}(\boldsymbol{q})\boldsymbol{M}_{ab}(\boldsymbol{q}),$$

is defined as the manipulator-to-base dynamic coupling matrix. Equation (4) describes how manipulator joint motion ($\dot{\boldsymbol{q}}_a$) generates reaction motion of the spacecraft base $\dot{\boldsymbol{q}}_b$ to preserve the total linear and angular momenta of the free-floating system. Joint motions belonging to the null space of $\boldsymbol{C}_{ba}$ produce no instantaneous base velocity, whereas motions outside this null space induce a coupled base response. However, in this study, we are interested in those configurations that produce base motion.

For further analysis, $\dot{\boldsymbol{q}}_b$ is split as

$$\dot{\boldsymbol{q}}_b = \begin{bmatrix} \boldsymbol{v}_b \\ \omega_b \end{bmatrix}, = \begin{bmatrix} \begin{bmatrix} \dot{x}_0 \\ \dot{y}_0 \end{bmatrix} \\ [\dot{q}_0] \end{bmatrix} == \boldsymbol{C}_{ba}(\boldsymbol{q})\dot{\boldsymbol{q}}_a = \begin{bmatrix} \boldsymbol{C}_{ba}^{v}(\boldsymbol{q}) \\ \boldsymbol{C}_{ba}^{\omega}(\boldsymbol{q}) \end{bmatrix} \dot{\boldsymbol{q}}_a = \begin{bmatrix} \boldsymbol{C}_{ba}^{v}(\boldsymbol{q})\dot{\boldsymbol{q}}_a \\ \boldsymbol{C}_{ba}^{\omega}(\boldsymbol{q})\dot{\boldsymbol{q}}_a \end{bmatrix}.$$

where $\boldsymbol{v}_b$ and $\omega_b$ denote base translation and rotation velocity. Similarly manipulator-to-base coupling matrix $\boldsymbol{C}_{ba}(\boldsymbol{q})$ is split as

$$\boldsymbol{C}_{ba}(\boldsymbol{q}) = \begin{bmatrix} \boldsymbol{C}_{ba}^{v}(\boldsymbol{q}) \\ \boldsymbol{C}_{ba}^{\omega}(\boldsymbol{q}) \end{bmatrix}$$

into $\boldsymbol{C}_{ba}^{v} \in \mathbb{R}^{2\times 3}$ and $\boldsymbol{C}_{ba}^{\omega} \in \mathbb{R}^{1\times 3}$. Consequently, base translation and rotation velocity are obtained as

$$\boldsymbol{v}_b = \boldsymbol{C}_{ba}^{v}\dot{\boldsymbol{q}}_a$$

and

$$\omega_b = \boldsymbol{C}_{ba}^{\omega}\dot{\boldsymbol{q}}_a.$$

That is, $\boldsymbol{C}_{ba}^{v}$ and $\boldsymbol{C}_{ba}^{\omega}$ isolate the effect of manipulator motion on base's linear and rotational velocity. Note the explicit dependence of $\boldsymbol{C}_{ba}^{v}$ and $\boldsymbol{C}_{ba}^{\omega}$ on $\boldsymbol{q}$ has been removed for neatness purposes.

### 3.2 Manipulator-EE coupling

Manipulator motion also determines the motion of the EE. From the differential kinematics introduced in Equation (3), the EE motion can be written as

$$\dot{\boldsymbol{q}}_e = \begin{bmatrix} \boldsymbol{v}_e \\ \omega_e \end{bmatrix} = \boldsymbol{J}_{e,b}\dot{\boldsymbol{q}}_b + \boldsymbol{J}_{e,a}\dot{\boldsymbol{q}}_a$$

Next, using Equation (4) yields

$$\dot{\boldsymbol{q}}_e = \boldsymbol{J}_{e,b}\boldsymbol{C}_{ba}(\boldsymbol{q})\dot{\boldsymbol{q}}_a + \boldsymbol{J}_{e,a}\dot{\boldsymbol{q}}_a = \left(\boldsymbol{J}_{e,b}\boldsymbol{C}_{ba} + \boldsymbol{J}_{ea}\right)\dot{\boldsymbol{q}}_a$$

$$\dot{\boldsymbol{q}}_e = \boldsymbol{C}_{ea}\dot{\boldsymbol{q}}_a, \tag{5}$$

where

$$\boldsymbol{C}_{ea} = \boldsymbol{J}_{eb}\boldsymbol{C}_{ba} + \boldsymbol{J}_{ea}$$

is defined as the manipulator-to-EE coupling matrix.

The matrices $\boldsymbol{C}_{ba}(\boldsymbol{q})$ and $\boldsymbol{C}_{ea}(\boldsymbol{q})$ therefore describe the two velocity relationships through which manipulator joint motion simultaneously generates spacecraft-base and EE motion.

### 3.3 EE to base coupling

The coupling matrices $\boldsymbol{C}_{ba}(\boldsymbol{q})$ and $\boldsymbol{C}_{ea}(\boldsymbol{q})$ describe how manipulator joint motion generate base and EE motion. In this study, the direct effect of EE motion on base motion is analyzed. Hence, a direct coupling matrix between the EE and the spacecraft base is necessary. From Equation (5), arm motion is written as

$$\dot{\boldsymbol{q}}_a = \boldsymbol{C}_{ea}^{\#}\dot{\boldsymbol{q}}_e,$$

where $(\cdot)^{\#}$ denotes the Moore-Penrose pseudoinverse. Substitution into Equation (4), the manipulator-to-base yields

$$\dot{\boldsymbol{q}}_b = \boldsymbol{C}_{ba}\boldsymbol{C}_{ea}^{\#}\dot{\boldsymbol{q}}_e = \boldsymbol{C}_{be}\dot{\boldsymbol{q}}_e, \tag{6}$$

where $\boldsymbol{C}_{be} = \boldsymbol{C}_{ba}\boldsymbol{C}_{ea}^{\#}$ is defined as the **EE-to-base dynamic coupling matrix**. Unlike $\boldsymbol{C}_{ba}$, which characterizes the base reaction generated by manipulator joint motion, $\boldsymbol{C}_{be}$ directly relates EE motion to the resulting spacecraft-base motion. It therefore provides a task-space representation of the dynamic interaction between the EE and the free-floating spacecraft.

### 3.4. Translational and Rotational EE-to-Base Coupling

The *coupling matrix* $\boldsymbol{C}_{be}$ relates the EE motion vector $\dot{\boldsymbol{q}}_e = [\boldsymbol{v}_e^T \quad \omega_e]^T$ to the full spacecraft-base motion vector $\dot{\boldsymbol{q}}_b = [\boldsymbol{v}_b^T \quad \omega_b]^T$, without independently identifying the translational and rotational motion-transfer mechanisms. These relationships are therefore examined separately using the corresponding translational and rotational components of $\boldsymbol{C}_{ba}$ as:

$$\dot{\boldsymbol{q}}_b = \boldsymbol{C}_{ba}\dot{\boldsymbol{q}}_a = \begin{bmatrix} \boldsymbol{C}_{ba}^{v} \\ \boldsymbol{C}_{ba}^{\omega} \end{bmatrix} \dot{\boldsymbol{q}}_a = \begin{bmatrix} \boldsymbol{C}_{ba}^{v}\dot{\boldsymbol{q}}_a \\ \boldsymbol{C}_{ba}^{\omega}\dot{\boldsymbol{q}}_a \end{bmatrix}. \tag{7}$$

where partition matrices $\boldsymbol{C}_{ba}^{v} \in \mathbb{R}^{2\times 3}$ and $\boldsymbol{C}_{ba}^{\omega} \in \mathbb{R}^{1\times 3}$ characterize the translational and rotational components of the arm-to-base coupling, respectively. Similarly,

$$\dot{\boldsymbol{q}}_e = \boldsymbol{C}_{ea}\dot{\boldsymbol{q}}_a = \begin{bmatrix} \boldsymbol{C}_{ea}^{v}\dot{\boldsymbol{q}}_a \\ \boldsymbol{C}_{ea}^{\omega}\dot{\boldsymbol{q}}_a \end{bmatrix}, \tag{8}$$

where $\boldsymbol{C}_{ea}^{v} \in \mathbb{R}^{2\times 3}$ and $\boldsymbol{C}_{ea}^{\omega} \in \mathbb{R}^{1\times 3}$ describe the corresponding arm-to-EE coupling components.

The isolated effects of EE translational ($\boldsymbol{v}_e$) and rotational ($\omega_e$) velocities on base's motion ($\boldsymbol{v}_b, \omega_b$) are obtained

$$\begin{aligned} \boldsymbol{v}_b &= \boldsymbol{C}_{ba}^{v}\dot{\boldsymbol{q}}_a = \boldsymbol{C}_{ba}^{v}(\boldsymbol{C}_{ea}^{v})^{\#}\boldsymbol{v}_e = \boldsymbol{C}_{be}^{vv}\boldsymbol{v}_e, \\ \omega_b &= \boldsymbol{C}_{ba}^{\omega}\dot{\boldsymbol{q}}_a = \boldsymbol{C}_{ba}^{\omega}(\boldsymbol{C}_{ea}^{v})^{\#}\boldsymbol{v}_e = \boldsymbol{C}_{be}^{\omega v}\boldsymbol{v}_e, \\ \boldsymbol{v}_b &= \boldsymbol{C}_{ba}^{v}\dot{\boldsymbol{q}}_a = \boldsymbol{C}_{ba}^{v}(\boldsymbol{C}_{ea}^{\omega})^{\#}\omega_e = \boldsymbol{C}_{be}^{v\omega}\omega_e, \\ \omega_b &= \boldsymbol{C}_{ba}^{\omega}\dot{\boldsymbol{q}}_a = \boldsymbol{C}_{ba}^{\omega}(\boldsymbol{C}_{ea}^{\omega})^{\#}\omega_e = \boldsymbol{C}_{be}^{\omega\omega}\omega_e. \end{aligned} \tag{9}$$

Here, $\boldsymbol{C}_{be}^{ij}$ characterizes EE translation/rotation to base translation/rotation. These mappings allow the physical nature of the EE-to-base interaction to be examined independently. They distinguish whether EE translation or rotation produces translational or rotational reaction motion of the spacecraft. This information complements the complete mapping $\boldsymbol{C}_{be}$, whose global coupling characteristics do not independently identify the underlying motion-transfer mechanism.

### 3.5 Singular Value Decomposition-Based Metrics

Several metrics have been proposed to quantify the dynamic coupling in FFRM. Xu et al. (1993) introduced the use of eigenvalue analysis to characterize the coupling magnitude and principal directions on the velocity mapping between the EE and the base. Later, Zhou et al. (2019) used **singular value decomposition** (SVD) of the coupling matrices to estimate coupling ellipsoids of the velocity. Peng et al. (2020) extended these metrics to dual-arm space robotic systems by defining generalized coupling factors to evaluate the multiple coupling effects among the manipulators and the spacecraft base. While these studies primarily focused on scalar measures of coupling strength, the present work emphasizes the modal characteristics of dynamic coupling. Specifically, SVD is employed to identify the strongest and weakest coupling modes, quantify the maximum velocity gain through the dominant singular value, and evaluate the alignment between the velocity directions of the base and EE. These metrics provide insight not only into the magnitude of coupling but also into the directional mechanisms through which motion is transmitted across the system. For a general coupling matrix $\boldsymbol{C}$ of rank $r$, the SVD is expressed as

$$\boldsymbol{C} = \boldsymbol{USV} = [\boldsymbol{U}_1 \quad \boldsymbol{U}_2] \begin{bmatrix} \boldsymbol{\Sigma}_1 & \boldsymbol{0} \\ \boldsymbol{0} & \boldsymbol{0} \end{bmatrix} \begin{bmatrix} \boldsymbol{V}_1^T \\ \boldsymbol{V}_2^T \end{bmatrix} = \boldsymbol{U}_1\, \boldsymbol{\Sigma}_1 \boldsymbol{V}_1^T, \tag{10}$$

where matrices $\boldsymbol{U}_1 = [\boldsymbol{u}_1 \quad \boldsymbol{u}_2 \quad \cdots \quad \boldsymbol{u}_r]$ and $\boldsymbol{V}_1 = [\boldsymbol{v}_1^T \quad \boldsymbol{v}_2^T \quad \cdots \quad \boldsymbol{v}_r^T]^T$ contain orthonormal left ($\boldsymbol{u}_i$) and right ($\boldsymbol{v}_i$) singular vectors, respectively, while matrix $\boldsymbol{\Sigma}_1 = diag\{\sigma_1, \sigma_2, \cdots \sigma_r\}$ contains the singular values satisfying $\sigma_1 \geq \sigma_2 \geq \cdots \geq \sigma_r > 0$. Matrix $\boldsymbol{C}$, of rank $r$, is written equivalently as

$$\boldsymbol{C} = \sum_{i=1}^{r} \sigma_i \, \boldsymbol{u}_i \boldsymbol{v}_i^T,$$

where each singular value satisfies

$$\boldsymbol{C}\boldsymbol{v}_i = \sigma_i \boldsymbol{u}_i$$

and has direct coupling interpretation. The right singular vector $\boldsymbol{v}_i$ identifies an input velocity direction, the singular value $\sigma_i$ quantifies the corresponding velocity gain, and the left singular vector $\boldsymbol{u}_i$ identifies the resulting output velocity direction. The modes associated with the largest and smallest singular values are subsequently referred to as the **strong** and **weak** coupling modes, respectively,

$$\boldsymbol{C}\boldsymbol{v}^s = \sigma_s \boldsymbol{u}^s, \boldsymbol{C}\boldsymbol{v}^w = \sigma_w \boldsymbol{u}^w.$$

### *3.5.1 Coupling strength and anisotropy:*

The maximum coupling strength at a given manipulator configuration is quantified through the dominant singular value. Accordingly, the **coupling strength** of a coupling matrix $\boldsymbol{C}$ is defined as

$$\sigma_s = \sigma_{\max}\,[\boldsymbol{C}(\boldsymbol{q})] = \sigma_1,$$

the maximum singular value of matrix $\boldsymbol{\Sigma}_1$ in the SVD. The coupling strength represents the maximum attainable velocity amplification between the corresponding input and output spaces at a given configuration. It identifies the strongest possible coupling mode but does not indicate how the coupling is distributed among the velocity directions. To quantify the directional dependence of the coupling, the **coupling anisotropy** of a coupling matrix $\boldsymbol{C}(\boldsymbol{q})$ is defined as

$$K_a(\boldsymbol{C}(\boldsymbol{q})) = \frac{\sigma_s}{\sigma_w} \tag{11}$$

where $\sigma_s$ and $\sigma_w$ denote the largest and smallest singular values of the mapping under consideration. Note $\sigma_s = \sigma_1$ and $\sigma_w = \sigma_r$. Values of $K_a$ close to unity indicate that the coupling gains are similar along the principal directions and therefore correspond to approximately isotropic coupling. Increasing values of $K_a$ indicate progressively stronger directional dependence, with the coupling becoming concentrated along the strong singular direction.

At rank-deficient configurations, the weak singular value may vanish. For $\sigma_w = 0$, $K_a$ is formally undefined and approaches infinity as $\sigma_w \to 0$. Such configurations represent the limiting case of completely directional coupling, for which motion along at least one output direction cannot be generated through the corresponding coupling relationship.

The coupling factor and anisotropy provide complementary information. The former characterizes the maximum attainable coupling magnitude, whereas the latter characterizes the relative difference between the strong and weak coupling modes.

## 3.6 Geometric and Directional Interpretation

### *3.6.1. Dynamic Coupling Ellipsoid*

SVD provides a geometric interpretation of dynamic coupling through coupling ellipsoids, which represent how a unit-norm set of input velocities is mapped into the corresponding output-velocity space.

For translational EE-to-base mapping

$$\boldsymbol{v}_b = \boldsymbol{C}_{be}^{vv}\boldsymbol{v}_e,$$

a unit circle of EE translational velocities is mapped into an ellipse of attainable base translational velocities. Its major and minor axes are oriented along the strong and weak base directions, respectively, and their lengths are proportional to $\sigma_s$ and $\sigma_w$. An approximately circular ellipse therefore corresponds to $K_{ij} \approx 1$, whereas an increasingly flattened ellipse corresponds to increasing anisotropy. If $\sigma_w = 0$, the mapping becomes rank deficient and the ellipse collapses into a line segment, indicating that the coupled base translation is restricted to a single direction.

The coupling ellipsoid thus provides a geometric representation of information contained in the singular values: its overall dimensions reflect the coupling magnitude, its eccentricity reflects the anisotropy, and its principal axes identify the directions of the resulting coupled motion.

### *3.6.2 Directional Velocity Alignment*

Although the coupling ellipsoid identifies the principal input and output directions, it does not by itself quantify the relative orientation between them. This relationship is particularly relevant for $\boldsymbol{C}_{be}^{vv}$, since an EE translational velocity along a principal input direction may produce a base velocity that is parallel, transverse, or opposite to the imposed EE motion.

For

$$\boldsymbol{C}_{be}^{vv} = \boldsymbol{U}_b \Sigma \boldsymbol{V}_e^T,$$

the singular-vector matrices are written as

$$\boldsymbol{U}_b = [\boldsymbol{u}_b^s \quad \boldsymbol{u}_b^w], \boldsymbol{V}_e = [\boldsymbol{v}_e^s \quad \boldsymbol{v}_e^w],$$

where $\boldsymbol{v}_e^s$ and $\boldsymbol{v}_e^w$ denote the strong and weak EE translational velocity directions, while $\boldsymbol{u}_b^s$ and $\boldsymbol{u}_b^w$ denote their corresponding base velocity directions.

The directional relationship between an EE velocity direction $\boldsymbol{v}_e^i$ and a base velocity direction $\boldsymbol{u}_b^i$ is quantified by the alignment angle

$$\alpha_{ij} = \cos^{-1}\left(\boldsymbol{v}_e^{i\,\mathrm{T}}\,\boldsymbol{u}_b^i\right),\ i, j \in \{s, w\}.$$

An alignment angle of $\alpha_{ij} = 0°$ indicates parallel EE and base velocity directions, $\alpha_{ij} = 90°$ indicates orthogonal directions, and $\alpha_{ij} = 180°$ indicates antiparallel directions. Four alignment angles are considered: $\alpha_{ss}$, $\alpha_{ww}$, $\alpha_{sw}$, and $\alpha_{ws}$, corresponding to the strong-to-strong, weak-to-weak, strong-to-weak, and weak-to-strong directional relationships, respectively.

The metrics introduced above provide complementary descriptions of dynamic coupling. The dominant singular value quantifies **how strongly** motion can be transmitted, the singular-value ratio quantifies **how directionally selective** that transmission is, and the alignment angle quantifies **where the resulting coupled motion is directed** relative to the imposed motion. These metrics form the basis for the configuration-dependent analysis presented in Section 4.

## 4. Characterization of EE-to-Base Coupling Dynamic

The dynamic coupling characterization is performed using the planar three-link FFRM model introduced in Section 2. The nominal geometric and inertial properties of the spacecraft base and manipulator are summarized in Table 1 and, unless otherwise stated, are used throughout the analysis. The dynamic coupling matrices are evaluated over the manipulator configuration space by varying $q_2$ and $q_3$, while the influence of $q_1$ is initially examined to assess the rotational symmetry of the coupling characteristics. At each configuration, the singular values and singular vectors of the coupling matrices are computed in MATLAB and used examine three complementary aspects of the coupling: its magnitude and anisotropy throughout the manipulator configuration space, the individual translational and rotational coupling relationships between the EE and the base, and the directional alignment between their coupled motions. Together, these analyses provide a characterization of not only how strongly EE motion affects the spacecraft base, but also how this interaction depends on the manipulator configuration and base mass, how directional the coupling is, and how the resulting base motion is oriented relative to the EE motion

*Table 1: Kinematic and dynamic parameters of the 3-link planar FFRM.*

| Parameter | Base | Link 1 | Link 2 | Link 3 |
|---|---|---|---|---|
| Mass $[kg]$ | $m_b = 20$ | $m_1 = 1$ | $m_2 = 2$ | $m_3 = 1$ |
| Inertia $[kg \cdot m^2]$ | $I_b = 0.1667$ | $I_1 = 0.0833$ | $I_2 = 0.6667$ | $I_3 = 0.0833$ |
| Length $[m]$ | $a = 0.3, b = 0.1$ | $l_1 = 1$ | $l_2 = 2$ | $l_3 = 1$ |
| CoM $[m]$ | $a_c = 0.15, b_c = 0.05$ | $l_{c1} = 0.5$ | $l_{c2} = 1$ | $l_{c3} = 0.5$ |

### 4.1 Configuration-Dependent Coupling Strength and Anisotropy

The first analysis examines how manipulator configuration and spacecraft base mass affect the strength and anisotropy of the dynamic coupling. Coupling maps are evaluated over the $(q_2, q_3)$ configuration space for the arm-to-base ($\boldsymbol{C}_{ba}$), arm-to-EE ($\boldsymbol{C}_{ea}$), and EE-to-base ($\boldsymbol{C}_{be}$) mappings. First, the influence of the first joint angle, ($q_1$), is examined. Next, the influence of the spacecraft base mass is investigated. Finally, the ratio between the strong and weak singular values is used to characterize the directional anisotropy of the EE-to-base coupling throughout the configuration space.

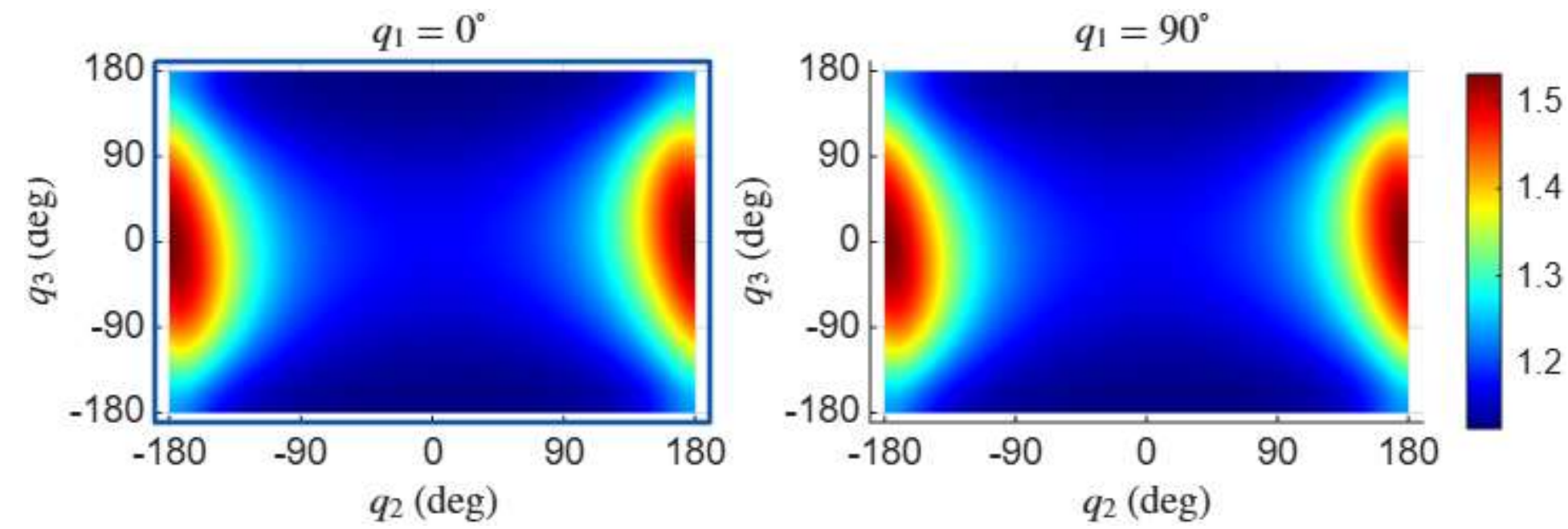

*Figure 2: Configuration-dependent base–arm coupling strength ($\sigma_s$) maps of the matrix $\boldsymbol{C}_{ba}$ over the ($q_2, q_3$) joint space for $q_1 = 0°$ (left) and $q_1 = 90°$ (right), with $m_b = 5m_a$.*

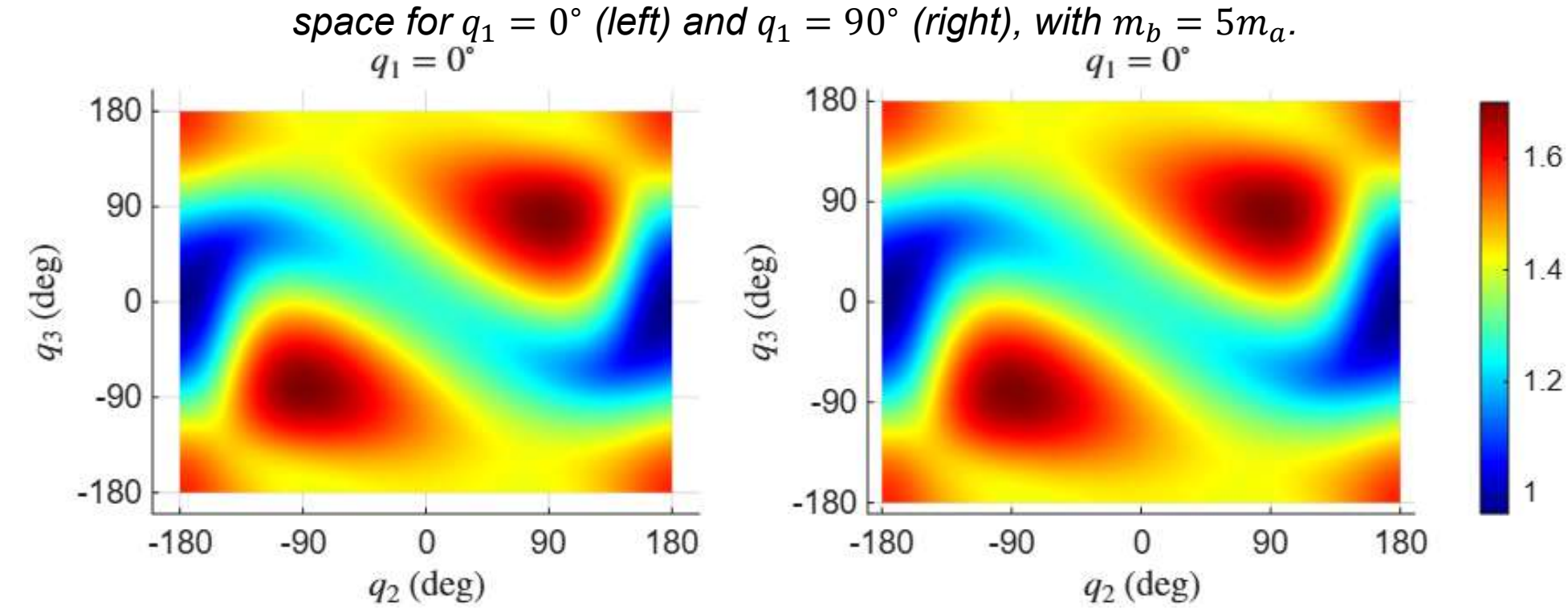

*Figure 3: Configuration-dependent EE–arm coupling maps of the matrix $\boldsymbol{C}_{ea}$ over the ($q_2, q_3$) joint space for $q_1 = 0°$ (left) and $q_1 = 90°$ (right), with $m_b = 5m_a$.*

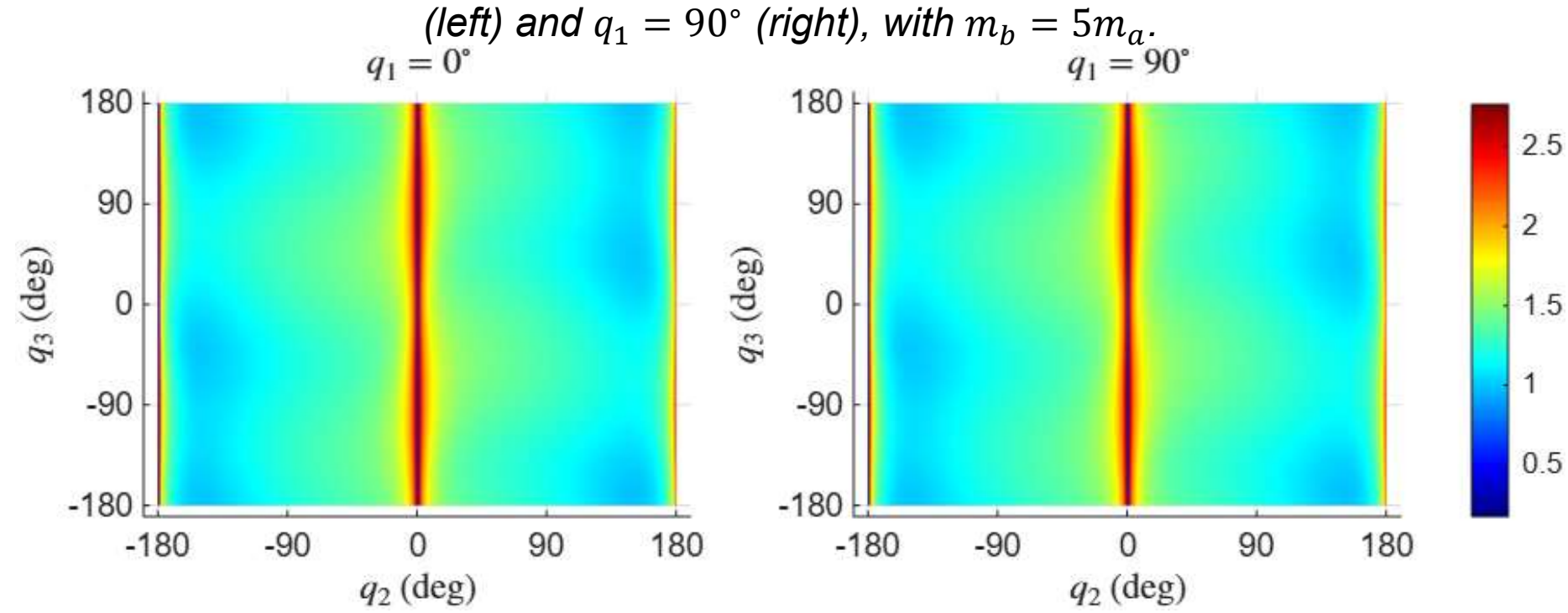

*Figure 4: Configuration-dependent EE–base coupling maps of the matrix $\boldsymbol{C}_{be}$ over the ($q_2, q_3$) joint space for $q_1 = 0°$ (left) and $q_1 = 90°$ (right), with $m_b = 5m_a$.*

The influence of the first joint angle on the dynamic coupling characteristics was first examined by evaluating the dominant singular values of the manipulator-to-base ($\boldsymbol{C}_{ba}$), manipulator -to-EE ($\boldsymbol{C}_{ea}$), and EE-to-base ($\boldsymbol{C}_{be}$) coupling matrices throughout the $(q_2, q_3)$ configuration space. Figures 2-4 compare the resulting coupling surfaces for $q_1 = 0°$ (left) and $q_1 = 90°$ (right), while maintaining a base-to-arm mass ratio of $m_b = 5m_a$. For all three mappings, the surfaces obtained at the two values of $q_1$ are identical: the locations of the maxima and minima, the overall surface topology, and the corresponding coupling magnitudes remain unchanged. This invariance results from the rotational symmetry of the planar system, since changing $q_1$ rotates the manipulator without modifying its relative geometry or inertial distribution. That is, $q_1$ changes the absolute directions of the base and EE velocity vectors in the inertial frame but not the coupling magnitudes. Therefore, $q_1$ is fixed at $0°$ in the remainder of the coupling-strength analysis, reducing the configuration space from three to two dimensions without loss of information regarding coupling magnitude. The individual maps nevertheless reveal markedly different dependencies on the relative arm configuration. For $\boldsymbol{C}_{ba}$ in Figure 2, the strongest manipulator-to-base coupling occurs primarily near $q_2 = \pm 180°$ and $q_3 \approx 0°$, indicating that these configurations maximize the ability of manipulator motion to induce base motion. In contrast, $\boldsymbol{C}_{ea}$ in Figure 3 exhibits its strongest coupling in configurations involving substantial relative rotations of the distal links, particularly around combinations of $q_2$ and $q_3$ near $\pm 90°$and toward the configuration-space boundaries, while reduced coupling is observed in regions near $q_3 = 0°$. The resulting EE-to-base mapping $\boldsymbol{C}_{be}$ in Figure 4 displays a distinctly different structure: its strongest coupling is concentrated around $q_2 = 0°$ and $q_2 = \pm 180°$, and the coupling generally decreases as $q_2$ moves away from these configurations toward intermediate flexion angles. Moreover, the $\boldsymbol{C}_{be}$ surface varies comparatively little with $q_3$, indicating that its dominant coupling strength is governed primarily by $q_2$. Thus, although $\boldsymbol{C}_{ba}$, $\boldsymbol{C}_{ea}$, and $\boldsymbol{C}_{be}$ describe related motion-transfer mechanisms, their configuration-dependent coupling landscapes are substantially different.

Having established that the coupling strength is invariant with respect to $q_1$, the first joint angle was fixed at $q_1 = 0°$ for the remainder of the coupling-strength analysis. The influence of the spacecraft base mass was then investigated by increasing the base-to-arm mass ratio from $m_b = 5m_a$ to $m_b = 20m_a$. Figure 5-7 compare the resulting coupling surfaces for $\boldsymbol{C}_{ba}$, $\boldsymbol{C}_{ea}$, and $\boldsymbol{C}_{be}$, respectively. In each case, the coupling strength is characterized by the largest singular value, $\sigma_1$, which represents the maximum velocity amplification attainable by the corresponding mapping at a given manipulator configuration. The results show that increasing the base mass affects the three coupling relationships differently: the manipulator-to-base coupling $\boldsymbol{C}_{ba}$ map is reduced, as expected for a more inertially dominant spacecraft; the manipulator -to-EE coupling $\boldsymbol{C}_{ea}$ map undergoes a comparatively smaller and strongly configuration-dependent redistribution; and the EE-to-base coupling $\boldsymbol{C}_{be}$ map exhibits the most pronounced reduction. Thus, the base mass not only changes the magnitude of the base response to manipulator motion but also modifies how the two intermediate mappings combine to determine the overall transmission of EE motion to the spacecraft.

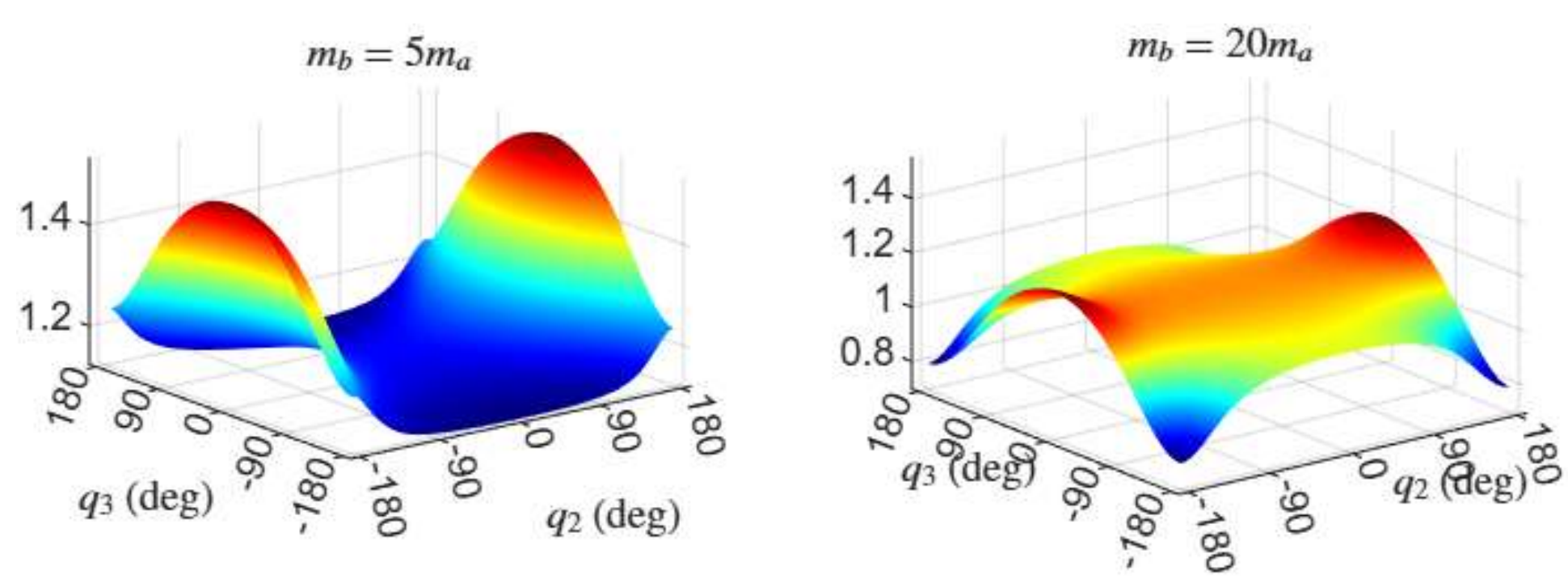

*Figure 5: Effect of the base-to-arm mass ratio on the coupling map of $\boldsymbol{C}_{ba}$ over the $(q_2, q_3)$ joint space, for $m_b = 5m_a$ (left) and $m_b = 20m_a$ (right), with $q_1 = 0°$.*

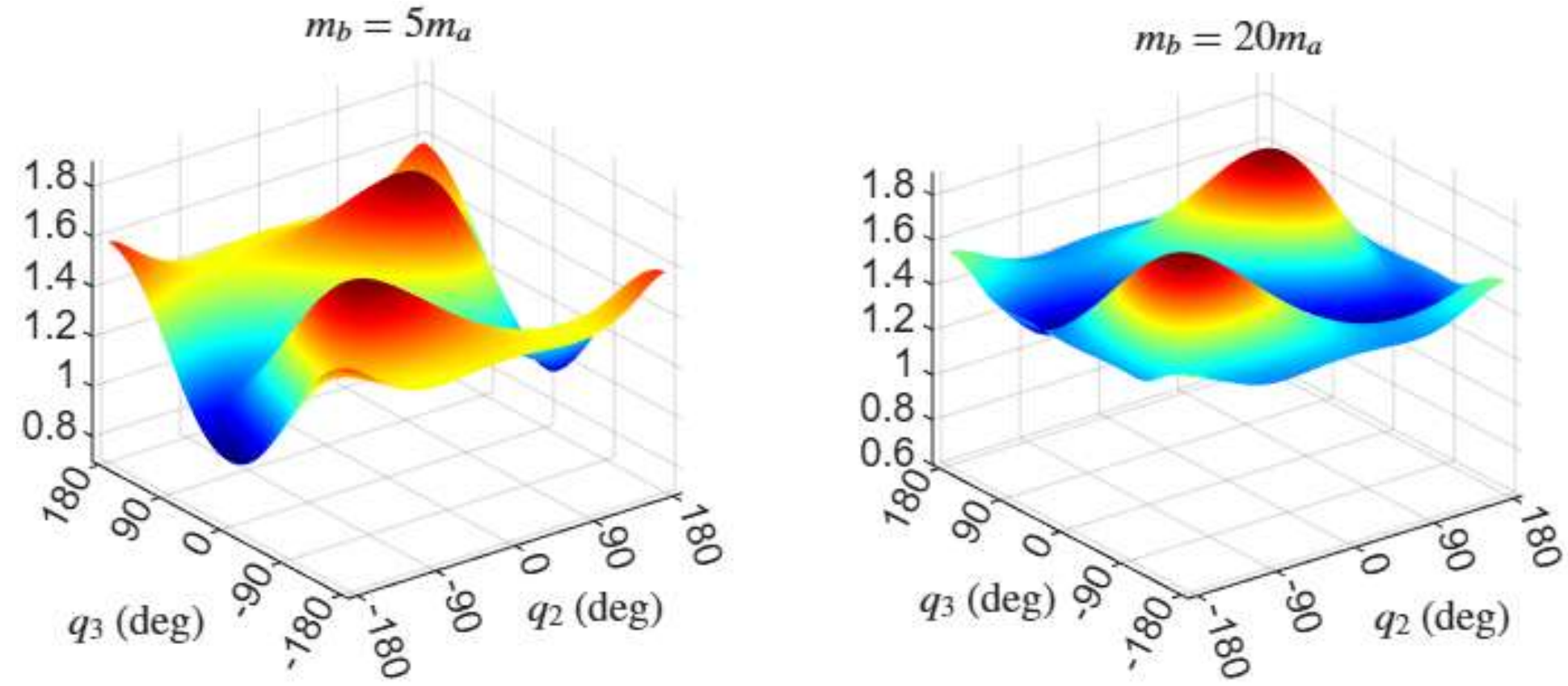

Figure 6: Effect of the base-to-arm mass ratio on the coupling map of $\boldsymbol{C}_{ea}$ over the $(q_2, q_3)$ joint space, for $m_b = 5m_a$ (left) and $m_b = 20m_a$ (right), with $q_1 = 0°$.

For the manipulator-to-base coupling mapping $\boldsymbol{C}_{ba}$ in Figure 5, increasing the base mass produces a clear reduction in the dominant coupling strength throughout the configuration space. However, the reduction is strongly configuration dependent rather than a uniform scaling of the original surface. For $m_b = 5m_a$, the largest coupling occurs near $q_2 = \pm 180°$ and $q_3 \approx 0°$, corresponding to configurations in which manipulator motion produces the strongest base response. When the base mass is increased to $m_b = 20m_a$, these high-coupling regions are substantially attenuated and the overall surface becomes flatter. The result confirms the expected inertial effect of a heavier spacecraft base while also demonstrating that its influence depends on manipulator configuration: increasing base mass provides the greatest attenuation precisely in configurations that exhibit the strongest joint-to-base coupling for the lighter spacecraft

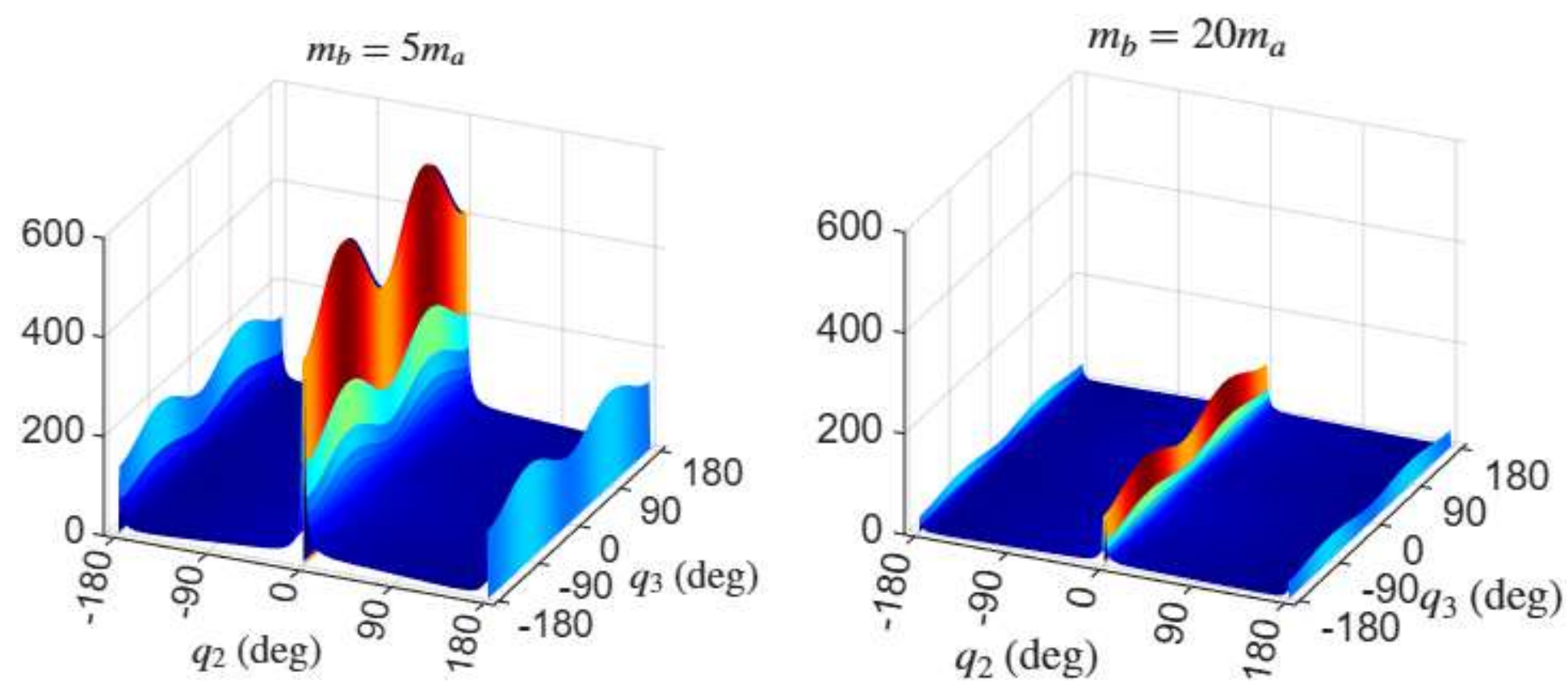

*Figure 7: Effect of the base-to-arm mass ratio on the coupling map of $\boldsymbol{C}_{be}$ over the $(q_2, q_3)$ joint space, for $m_b = 5m_a$ (left) and $m_b = 20m_a$ (right), with $q_1 = 0°$.*

The effect of base mass on the joint-to-EE mapping for matrix $\boldsymbol{C}_{ea}$ in Figure 6 is qualitatively different. The strongest coupling remains concentrated near the bent configurations around $q_2 \approx q_3 \approx \pm 90°$, surprisingly, its maximum value increases as the base mass is increased, contrary to dictated by intuition. Similarly, configurations that exhibit comparatively weak coupling for $m_b = 5m_a$, particularly those near $q_2 = \pm 180°$ and $q_3 \approx 0°$, experience an increase in coupling strength for $m_b = 20m_a$. Consequently, increasing the base mass does not attenuate $\boldsymbol{C}_{ea}$ but increases it and tends to homogenize the coupling strength throughout the configuration space. Physically, as the base becomes more inertially dominant, its reaction motion decreases and the manipulator-to-EE mapping progressively approaches the behavior of a manipulator supported by a less mobile base. That is, for a heavier base, the manipulator motion has a stronger effect on the EE compared to a lighter base.

The influence of base mass becomes considerably more pronounced for the complete EE-to-base mapping $\boldsymbol{C}_{be}$ in Figure 7. For both mass ratios, the coupling surface is characterized by a dominant ridge around $q_2 = 0°$, secondary high-coupling regions near $q_2 = \pm 180°$, and broad low-coupling plateaus at intermediate values of $q_2$. These structures extend almost uniformly along $q_3$, further demonstrating the weak dependence of the dominant $\boldsymbol{C}_{be}$ coupling on the third joint angle. Most importantly, increasing the base mass from $5m_a$ to $20m_a$ produces a substantial reduction of the entire coupling surface while preserving its characteristic topology. Therefore, base mass primarily scales the magnitude of the EE-to-base transmission, whereas the locations of strongly and weakly coupled configurations remain predominantly determined by manipulator geometry. Compared with the more moderate changes observed independently in $\boldsymbol{C}_{ba}$ and $\boldsymbol{C}_{ea}$, the considerably stronger mass sensitivity of $\boldsymbol{C}_{be}$ demonstrates that the overall transmission from EE motion to base motion emerges from the combined behavior of both constituent mappings and cannot be inferred from either one in isolation

The complete EE-to-base mapping, $\boldsymbol{C}_{be}$, provides a global characterization of the coupling between the full EE vector ($\dot{\boldsymbol{q}}_e = [\boldsymbol{v}_e^T \; \omega_e]^T$) and base velocity vector ($\dot{\boldsymbol{q}}_b = \left[\boldsymbol{v}_b^T \; \omega_b\right]^T$). However, its singular values combine translational and rotational motion into a single coupling measure and therefore do not identify the physical motion components responsible for the resulting base response. In particular, a large coupling gain in $\boldsymbol{C}_{be}$ does not distinguish whether the dominant interaction corresponds to EE translation inducing base translation or base rotation, or either of the analogous responses produced by EE rotation. To resolve these individual motion-transfer mechanisms, the translational and rotational relationships

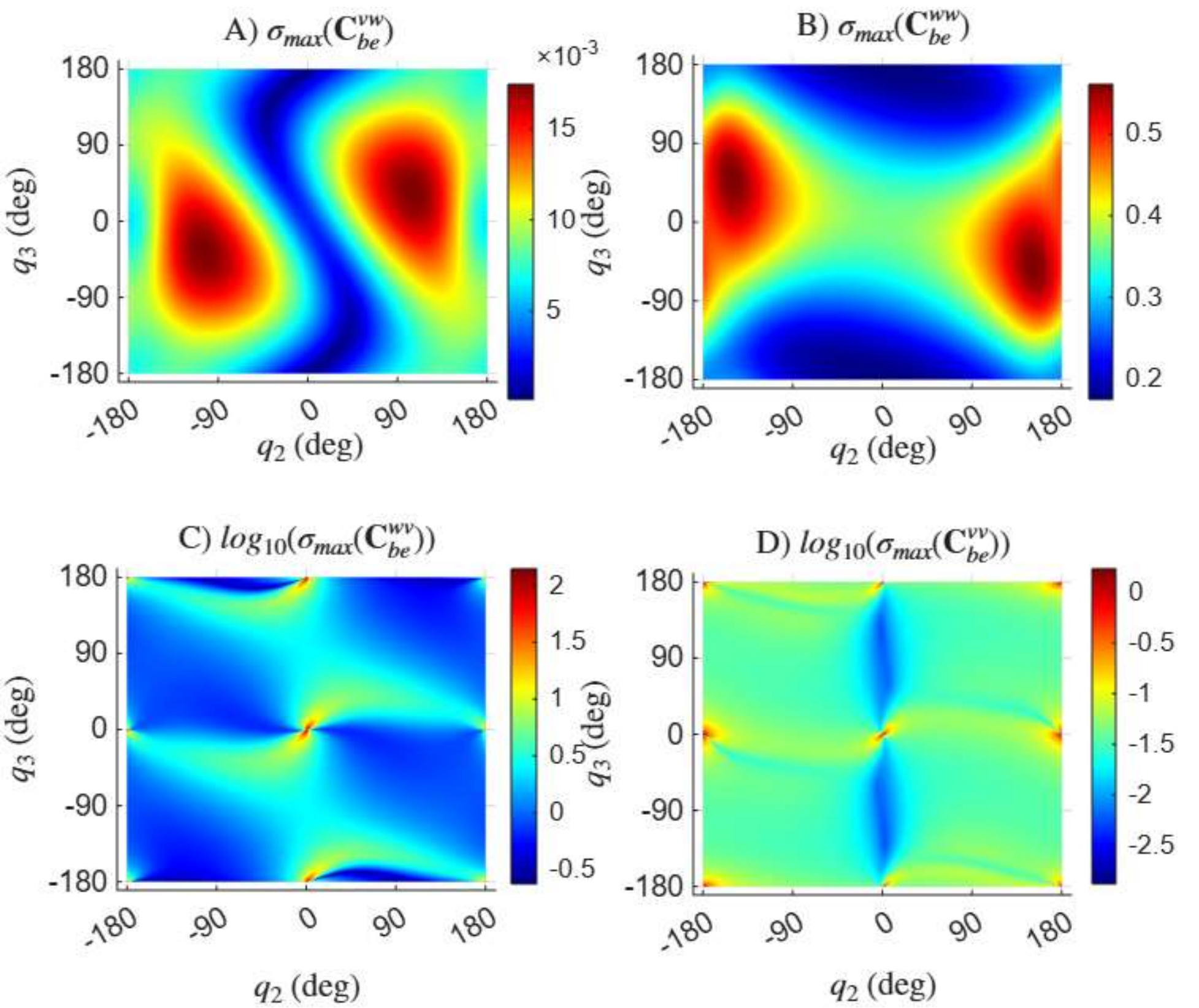

*Figure 8: Coupling-factor maps over the $(q_2, q_3)$ joint space. Panels (A) and (B) show the coupling gains associated with coupling matrices ($\boldsymbol{C}_{be}^{vw}$ ) and ($\boldsymbol{C}_{be}^{ww}$), respectively, while panels (C) and (D) present logarithmic representations of $\boldsymbol{C}_{be}^{wv}$ and $\boldsymbol{C}_{be}^{vv}$, highlighting regions of strong and weak coupling and sharp variations near singular configurations.*

are examined independently through their corresponding velocity-level coupling mappings.

To further characterize these motion-coupling mechanisms, the four velocity-level coupling matrices ($\boldsymbol{C}_{be}^{vv}$, $\boldsymbol{C}_{be}^{\omega v}$, $\boldsymbol{C}_{be}^{v\omega}$ and $\boldsymbol{C}_{be}^{\omega\omega}$) were independently evaluated throughout the $(q_2, q_3)$ configuration space. Figure 8 presents the maximum singular value of each mapping, thereby identifying the maximum attainable coupling gain for the corresponding input-output relationship at every manipulator configuration. The results reveal substantially different configuration-space structures for the four coupling mechanisms. The rotation-to-translation mapping $\boldsymbol{C}_{be}^{v\omega}$ is shown in Figure 8(A), which exhibits two localized regions of strong coupling approximately centered around $q_2 = \pm 100°$, with corresponding intermediate values of $q_3$ about $\pm 35$, separated by a pronounced diagonal low-coupling region. Thus, the extent to which EE rotation induces base translation is strongly dependent on the relative manipulator geometry. In contrast, the rotation-to-rotation mapping $\boldsymbol{C}_{be}^{\omega\omega}$ in Figure 8(B) exhibits stronger coupling toward $q_2 = \pm 150°$ and considerably weaker coupling over broad regions near $q_2$ between -$90°$and $90°$, particularly as $q_3$ increases. Hence, configurations that strongly transmit EE rotation into base translation do not necessarily maximize the corresponding rotational response of the base.

A similarly important distinction is observed when the imposed EE motion is translational. The translation-to-rotation mapping $\boldsymbol{C}_{be}^{\omega v}$ in Figure **8**(C) shows a strongly configuration-dependent response, with the coupling generally increasing with $q_2$ over substantial portions of the configuration space and exhibiting narrow high-gain regions near boundary configurations. By contrast, the translation-to-translation mapping $\boldsymbol{C}_{be}^{vv}$ presented in Figure **8**(D) contains pronounced low-coupling valleys around $q_2 \approx 0°$, indicating configurations in which the translational motion induced at the base by EE translation can be substantially reduced. Importantly, corresponding minima are not generally present in $\boldsymbol{C}_{be}^{\omega v}$. Therefore, selecting a configuration that suppresses base translation does not necessarily suppress the base rotation generated by the same EE translation motion. Conversely, configurations away from these low $\boldsymbol{C}_{be}^{vv}$ regions may increase the translational response while producing a different rotational sensitivity. This distinction demonstrates that the effect of a prescribed type of EE motion cannot be characterized solely by the magnitude of the complete $\boldsymbol{C}_{be}$ mapping, since the translational and rotational components of the resulting base motion may exhibit different—and in some regions opposing—configuration-dependent trends.

Taken together, these results show that the same manipulator configuration may favor one motion-transmission mechanism while suppressing another. The four mappings therefore provide complementary information regarding how EE translation and rotation are individually transmitted into base translation and rotation. In particular, a configuration that appears favorable when only base translation is considered may remain unfavorable with respect to base rotation, emphasizing the importance of identifying the specific base response of interest when evaluating or selecting manipulator configurations.

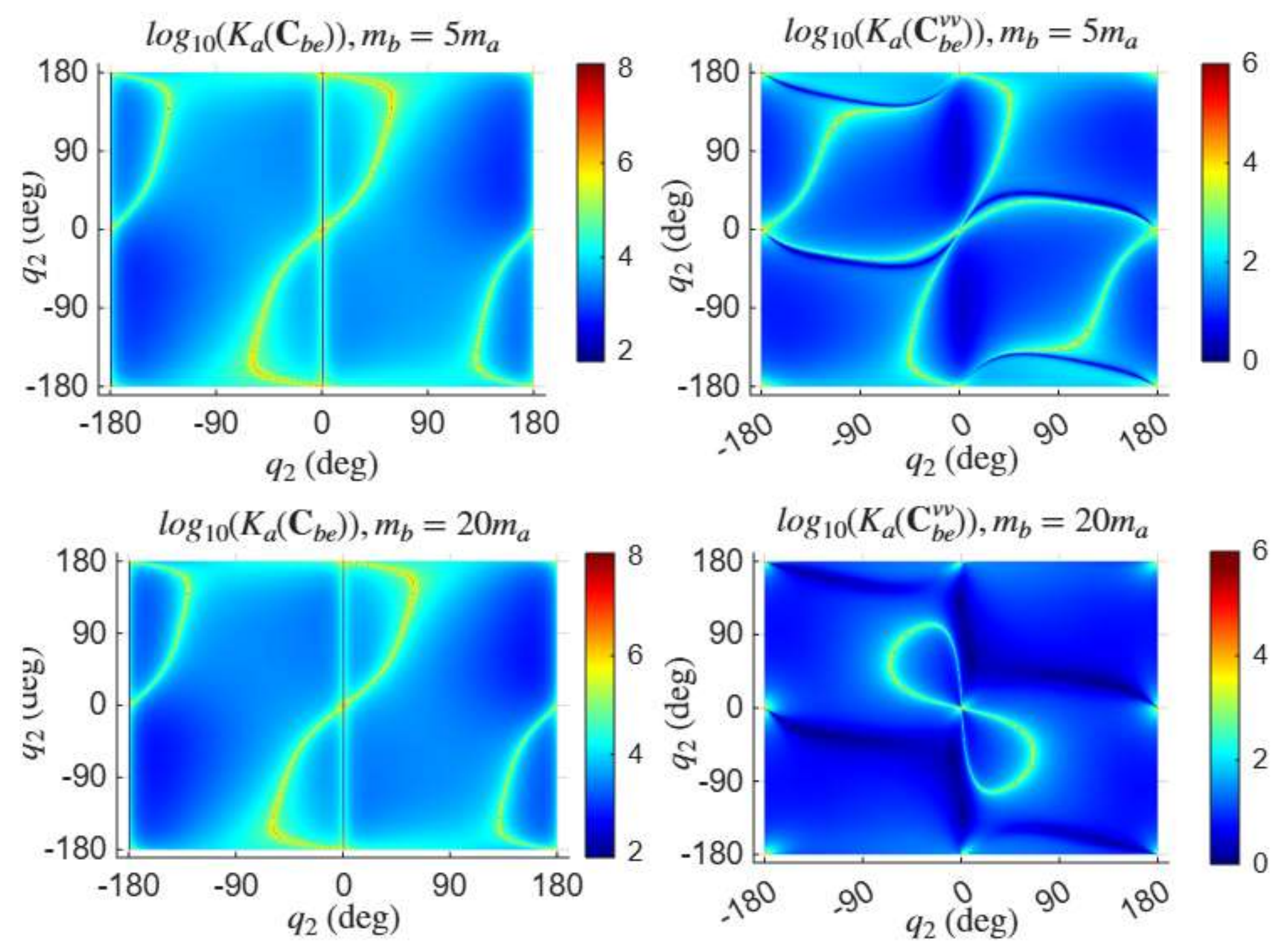

*Figure 9: Maps of the coupling anisotropy factor, ($log_{10}(K_{ij})$ ), over the ($q_2, q_3$) joint space. The left panels (A, C) correspond to the complete end-effector-to-base velocity mapping ($\boldsymbol{C}_{be}$), while the right panels (B, D) represent its translational submapping ($\boldsymbol{C}_{be}^{vv}$). The upper and lower panels correspond to ($m_b = 5m_a$) and ($m_b = 20m_a$), respectively. Higher values indicate stronger directional dependence of the velocity coupling.*

While the largest singular value used in the preceding analyses characterizes the maximum coupling strength, it does not indicate how strongly the response depends on the direction of the imposed velocity. The directional analysis is focused on the EE-to-base relationship, $\boldsymbol{C}_{be}$, since this mapping directly characterizes the influence of EE motion on the spacecraft base, and constitutes the primary coupling relationship of interest in this study. The translational mapping $\boldsymbol{C}_{be}^{vv}$ is additionally considered to isolate the directional characteristics of the linear velocity coupling and to provide a direct geometric interpretation through the velocity-coupling ellipses to be presented in Section 4.2. To quantify this directional dependence, the coupling anisotropy $K_{ij}$ is defined in Equation (11), and $\log_{10}(K_{ij})$ is mapped throughout the $(q_2, q_3)$ configuration space. Figure 9 presents this metric for the complete EE-to-base mapping $\boldsymbol{C}_{be}$ on the left (A and C panels) and for the translational sub mapping $\boldsymbol{C}_{be}^{vv}$ on the right (B and D); for both $m_b = 5m_a$ (A and B) and $m_b = 20m_a$ (C and D). The maps reveal that the coupling is strongly anisotropic and highly configuration dependent, with narrow regions of very large $K_{ij}$ indicating configurations in which EE motion is transmitted strongly to the base along one direction but only weakly along another. Interestingly, the anisotropy landscape of the complete $\boldsymbol{C}_{be}$ mapping in left plots is nearly unchanged when the base mass is increased, despite the substantial reduction in coupling magnitude observed previously. This indicates that increasing the base mass primarily attenuates the overall coupling strength without fundamentally altering its directional structure. The translational mapping $\boldsymbol{C}_{be}^{vv}$ shown in right-plots, however, exhibits a more pronounced dependence on base mass, with noticeable changes in both the magnitude and distribution of its anisotropic regions. Therefore, coupling strength and coupling anisotropy represent distinct characteristics of the free-floating system: a configuration with relatively small $\sigma_1$ may still be highly directional if $\sigma_{\min}$ is much smaller, while a strongly coupled configuration need not necessarily be highly anisotropic.

### 4.2 Geometric and Directional Characteristics of EE-to-Base Coupling

The coupling-strength and anisotropy maps presented in Section 4.1 quantify the magnitude and directional dependence of the EE-to-base interaction throughout the manipulator configuration space. However, these scalar measures do not by themselves reveal the physical directions of the EE motions associated with the strong and weak coupling modes, nor the directions of the resulting base motion. This section therefore examines the geometric information contained in the singular vectors of the translational EE-to-base coupling matrix $\boldsymbol{C}_{be}^{vv}$. The analysis present in this section first illustrates the coupling geometry using representative manipulator configurations, then extends this interpretation throughout the $(q_2, q_3)$configuration space, and finally quantifies the directional alignment between the principal EE and base velocity directions.

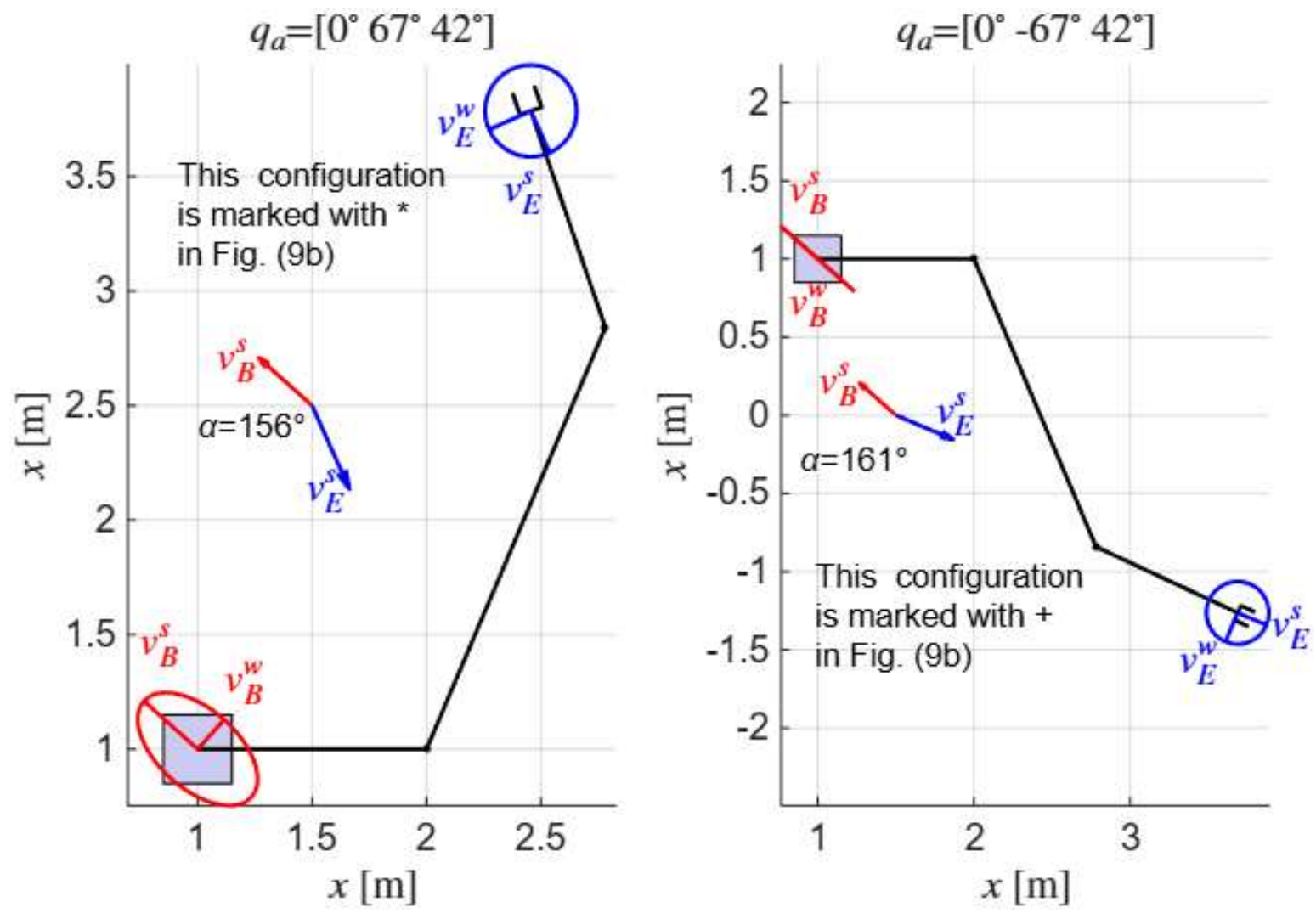

Figure 10: Representative FFRM configurations illustrating the directional dependence of EE-to-base velocity coupling for $q_2 = 67°$ (left) and $q_2 = -67°$ (right). Blue circles represent EE velocity directions, while red ellipses show the corresponding base velocity response. Changing $q_2$ produces a nearly collapsed base velocity ellipse, indicating strong coupling anisotropy and a pronounced directional dependence of the base reaction. For the translational mapping $\boldsymbol{v}_b = \boldsymbol{C}_{be}^{vv}\boldsymbol{v}_e$, the singular value decomposition is expressed as $\boldsymbol{C}_{be}^{vv} = \boldsymbol{U\Sigma V}^T$, where

$$\boldsymbol{\Sigma} = \begin{bmatrix} \sigma_s & 0 \\ 0 & \sigma_w \end{bmatrix},\ \boldsymbol{V} = [\ \boldsymbol{v}_e^s\ \boldsymbol{v}_e^w\ ],\ \boldsymbol{U} = [\ \boldsymbol{u}_b^s\ \boldsymbol{u}_b^w\ ].$$

The singular vectors $\boldsymbol{v}_e^s$ and $\boldsymbol{v}_e^w$ define the strong and weak EE translational velocity directions, whereas the corresponding left singular vectors $\boldsymbol{u}_b^s$ and $\boldsymbol{u}_b^w$ define the directions of the induced base translation. Geometrically, $\boldsymbol{C}_{be}^{vv}$ maps a unit circle of EE translational velocities into an ellipse of base velocities. The major and minor axes of this ellipse are oriented along $\boldsymbol{v}_B^s$ and $\boldsymbol{v}_B^w$, with lengths proportional to $\sigma_s$ and $\sigma_w$, respectively. Thus, the ellipse provides a direct geometric representation of both coupling strength and anisotropy.

Figure 10 illustrates this geometric interpretation for two representative manipulator configurations, $\boldsymbol{q}_a = [0°, 67°, 42°]^T$and $\boldsymbol{q}_a = [0°, -, 67°42°]^T$. The strong and weak EE velocity directions are shown at the EE, while the corresponding ellipse at the spacecraft base represents the attainable translational base velocities. The figure demonstrates that changing the manipulator configuration modifies both the shape and orientation of the coupling ellipse and therefore changes how EE translation is transmitted to the base. In particular, EE motion along $\boldsymbol{v}_e^s$ generates the maximum base translational response along $\boldsymbol{v}_B^s$, whereas motion along $\boldsymbol{v}_E^w$ produces the minimum response along $\boldsymbol{v}_B^w$. For the two configurations shown, the ellipse at the base collapses for the configuration shown in the right figure. Importantly, the input and output principal directions are not generally aligned, the angles between $\boldsymbol{v}_E^s$ and $\boldsymbol{v}_B^s$ are approximately $156°$and $161°$, indicating that the strongest base response is predominantly opposite to the corresponding EE motion. These examples therefore show that configuration determines not only how strongly EE motion is transmitted to the base, but also the direction into which that motion is redirected.

The two examples in Figure 10 provide a local interpretation of the coupling geometry. To determine whether these characteristics persist or change throughout the manipulator workspace, the same SVD-based representation was evaluated at representative configurations spanning the complete $(q_2, q_3)$ space.

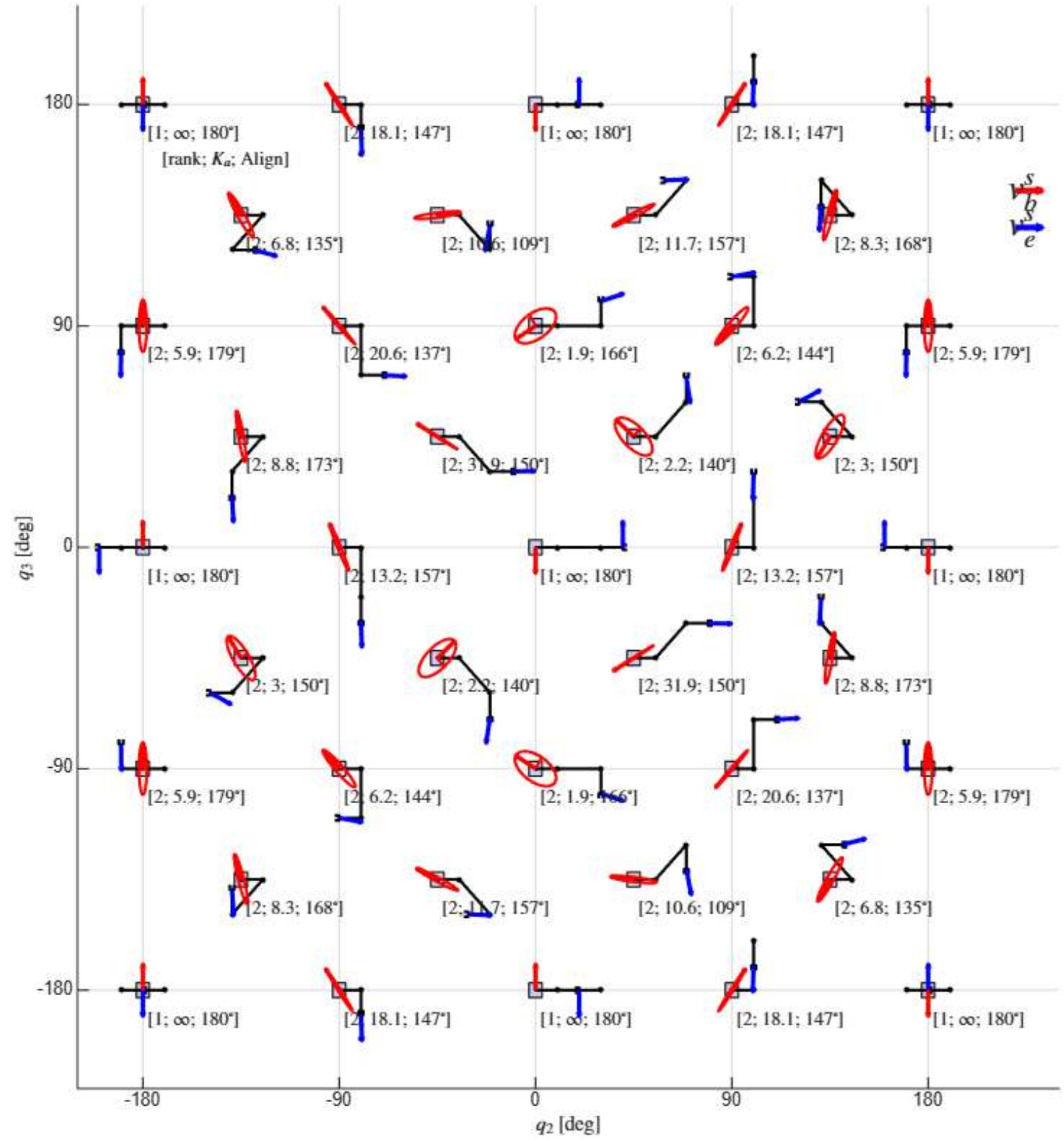

*Figure 11: Representative FFRM configurations and corresponding velocity-coupling ellipses distributed over the (q_2)–(q_3) joint space. Blue arrows indicate end-effector velocity vectors along the directions of strongest coupling, while red ellipses represent the corresponding induced base velocity, with nearly collapsed ellipses indicating strong coupling anisotropy.*

Figure 11 shows the evolution of the translational coupling geometry throughout the configuration space. Substantial variations are observed in the orientation and eccentricity of the base-velocity ellipses, demonstrating that the directional structure of $\boldsymbol{C}_{be}^{vv}$ is strongly configuration dependent. Configurations with $K_{ij} = \sigma_s/\sigma_w$ close to unity produce nearly circular ellipses at the base, indicating that the coupling magnitude is relatively insensitive to the direction of EE translation. As $K_{ij}$ increases, the ellipse becomes progressively flatter, indicating that the base response becomes concentrated along $\boldsymbol{v}_b^s$ while the response along $\boldsymbol{v}_b^w$ is increasingly suppressed. As expected, highly anisotropic configurations possess a preferred direction through which EE translation is transmitted to the base.

Of particular interest are the configurations for which $\mathrm{rank}(\boldsymbol{C}_{be}^{vv}) = 1$, visible at several configurations in Figure 11 (for instance, see $q_2 = q_3 = -180°$). At these configurations, the weak singular value vanishes, $\sigma_w = 0$, and the velocity ellipse collapses into a line segment. Consequently, EE translation can induce base translation only along a single direction, while the orthogonal base-velocity direction cannot be generated through this coupling. Since $K_{ij} = \sigma_s/\sigma_w$, the anisotropy ratio is formally undefined at these rank-one configurations and approaches infinity as $\sigma_w \to 0$, representing the limiting case of completely directional coupling. The figure also reveals substantial changes in the orientation of $\boldsymbol{v}_e^s$ and $\boldsymbol{v}_b^s$, with their alignment angle ranging from nearly antiparallel to considerably smaller angular separations depending on configuration.

The preceding geometric analysis therefore reveals two distinct directional properties of the coupling. The singular-value ratio $K_{ij}$ describes how strongly the coupling depends on the EE motion direction, whereas the relative orientation of the EE and base singular vectors describes where the resulting base motion is directed. To characterize the latter systematically, the alignment between the principal EE and base velocity directions is evaluated throughout the complete configuration space. For each pair of singular directions, the alignment angle is defined as $\alpha_{ij} = \cos^{-1}\left(\boldsymbol{v}_e^{i\,T}\boldsymbol{v}_b^j\right)$, $i, j \in \{s, w\}$, where the singular vectors are unit vectors. Accordingly, $\alpha_{ij} = 0°,\ 90°, 180°$ represent parallel, orthogonal, and antiparallel motion.

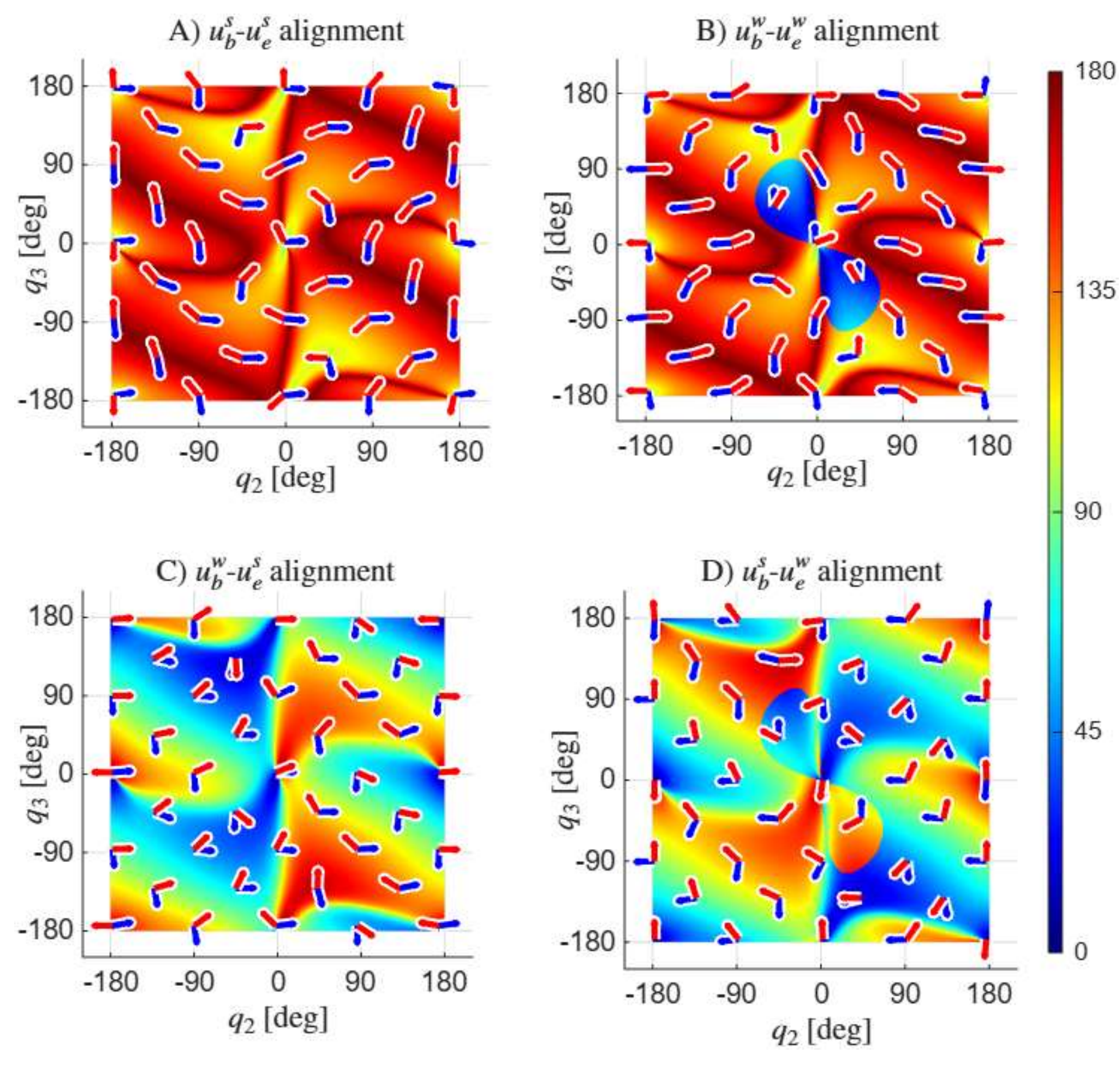

*Figure 12: Alignment maps between the strong and weak end-effector (blue arrows) and base (red arrows) velocity directions. Panels (A–D) show the angular alignment between strong–strong, weak–weak, weak–strong, and strong–weak directions, respectively. Color contours indicate the alignment angle (0°–180°), revealing configurations where the corresponding velocity directions are aligned, orthogonal, or opposed.*

Figure 12 maps the strong alignment angles between EE and base velocity vectors throughout the $(q_2, q_3)$ configuration space. The strong-to-strong alignment in is dominated by large angles over most of the domain, confirming that the behavior illustrated by the representative configurations in Figure 12(A) is not isolated. Instead, EE translation along the strongest coupling direction generally induces a base velocity directed predominantly opposite the EE motion, providing a geometric representation of the characteristic reaction or push-back behavior of the free-floating system. A similar predominantly opposing relationship is observed in Figure 12(B) for the weak-to-weak directions over much of the configuration space. However, this map also contains a distinct central, hourglass-shaped region in which the alignment angle becomes substantially smaller. Within this region, the weak EE and base velocity directions transition from predominantly opposing to predominantly codirectional motion. The relatively sharp boundaries separating these regimes demonstrate that the directional relationship between EE and base motion can change significantly with relatively small variations in manipulator configuration. These boundaries should nevertheless be interpreted with consideration of the sign and mode-selection properties inherent to the SVD, particularly near configurations where singular values approach one another.

The cross-alignment maps, corresponding to strong EE–weak base and weak EE–strong base directions in Figure 12(C-D), exhibit complementary spatial patterns. This behavior is consistent with the orthogonality of the strong and weak singular-vector bases, since $\boldsymbol{v}_E^s \perp \boldsymbol{v}_E^w$ and $\boldsymbol{v}_B^s \perp \boldsymbol{v}_B^w$. Consequently, the four alignment relationships are not independent but collectively describe the relative orientation between the EE and base principal velocity frames.

Taken together, Figure 12(A-D) demonstrates that the EE-to-base translational coupling possesses three complementary characteristics. The singular values determine the **coupling strength**, their ratio determines the **coupling anisotropy**, and the relative orientation of the corresponding singular vectors determines the **directional alignment** of the coupled motion. Manipulator configuration therefore determines not only how much base motion can be induced by EE translation, but also how selectively that motion is transmitted with respect to direction and whether the resulting base response is predominantly codirectional, transverse, or opposite to the EE motion. These characteristics provide information that cannot be obtained from a single scalar coupling factor and form a more complete geometric description of the configuration-dependent dynamic coupling in the free-floating manipulator.

## 5. Conclusions and Future Work

This study presented a systematic characterization of dynamic coupling in a planar three-link FFRM, considering the magnitude of the coupling, its anisotropy, and its directional characteristics. SVD was used to identify the strong and weak coupling modes throughout the manipulator configuration space and to provide both quantitative and geometric interpretations of the motion transmitted between the manipulator, EE, and spacecraft base. The results showed that the coupling strength is invariant with respect to the first joint angle $q_1$, consistent with the rotational symmetry of the planar system about $q_1$, although $q_1$can modify the absolute orientation of the associated coupling directions. In contrast, the relative configuration defined by $q_2$ and $q_3$ strongly influences the coupling characteristics and determines the locations of strongly and weakly coupled configurations.

The influence of spacecraft inertia was investigated by increasing the base-to-manipulator mass ratio from $m_b = 5m_a$ to $m_b = 20m_a$. Increasing the base mass reduced the joint-to-base coupling and produced an even more pronounced attenuation of the EE-to-base coupling, whereas its influence on the manipulator-to-EE mapping was comparatively smaller and more configuration dependent. Importantly, changes in base mass did not affect all coupling characteristics in the same manner. Substantial reductions in coupling magnitude could occur while the overall anisotropy pattern remained comparatively unchanged, demonstrating that coupling strength and directional structure constitute distinct properties of the dynamic interaction.

The analysis of the individual EE-to-base velocity mappings further showed that translational and rotational motions are transmitted through distinct configuration-dependent mechanisms. The mappings $\boldsymbol{C}_{be}^{vv}$, $\boldsymbol{C}_{be}^{\omega v}$, $\boldsymbol{C}_{be}^{v\omega}$, and $\boldsymbol{C}_{be}^{\omega\omega}$ revealed that a configuration that attenuates one type of base response does not necessarily suppress another. Configurations exhibiting reduced base translation in response to EE translation may still produce appreciable base rotation. Thus, the overall EE-to-base coupling cannot completely describe the physical nature of the induced spacecraft motion without distinguishing its translational and rotational components.

The SVD-based geometric analysis showed that the translational EE-to-base coupling can also be strongly anisotropic. The ratio between the strong and weak singular values identifies configurations in which EE motion is transmitted very differently depending on its direction. As the weak singular value approaches zero, the corresponding base-velocity ellipse becomes increasingly flattened, and at rank-one configurations it collapses into a line segment, restricting the coupled base translation to a single direction. The anisotropy ratio consequently tends to infinite as the weak singular value approaches zero, representing the limiting case of completely directional coupling.

Finally, analysis of the singular vectors demonstrated that coupling magnitude and anisotropy alone do not completely characterize the interaction. The orientation of the principal EE and base velocity directions varies substantially throughout the configuration space. Large regions exhibit a characteristic push-back behavior, in which the induced base translation is predominantly opposite to the corresponding EE motion, whereas other configurations produce transverse or approximately codirectional responses. Taken together, these results indicate that dynamic coupling is characterized by three complementary properties: **coupling strength**, which determines how much motion can be transmitted; **coupling anisotropy**, which determines how strongly that transmission depends on the motion direction; and **directional alignment**, which determines where the resulting base motion is directed relative to the EE motion.

Future work will extend this intrinsic characterization to the coupling produced by **prescribed task-space motions**. The present SVD analysis identifies the EE directions associated with the maximum and minimum attainable coupling at each configuration; however, during an actual manipulation task, the required EE velocity direction is imposed by the task and will generally not coincide with either singular direction. For a prescribed translational direction $\hat{\boldsymbol{d}}_e$, the resulting base response can be evaluated through $\boldsymbol{v}_b = C_{be}^{vv}\hat{\boldsymbol{d}}_e$, allowing the realized coupling gain and induced base-motion direction to be mapped throughout the configuration space. This extension will investigate how the orientation of a prescribed task relative to the strong and weak coupling directions determines the actual spacecraft response, including whether configurations with large maximum coupling can nevertheless produce small base disturbances when the required EE motion is favorably oriented. Such an analysis can identify dynamically favorable configurations for specific manipulation directions without requiring full trajectory optimization. The framework can subsequently be extended to prescribed rotational EE motions and the remaining translational–rotational coupling relationships, providing a natural intermediate step toward coupling-informed motion planning and control of FFRMs.